\documentclass[sigconf]{acmart}

\renewcommand\footnotetextcopyrightpermission[1]{}
\usepackage{amsmath,amsfonts,bm}
\usepackage{mathtools}
\usepackage{booktabs}
\usepackage{multirow}
\usepackage{array}
\usepackage{tabularx}
\usepackage{makecell}
\usepackage{enumitem}
\usepackage{xcolor} 
\usepackage{graphicx}
\usepackage{algorithm}
\usepackage{algpseudocode}
\usepackage{tikz}
\usepackage{subcaption}
\usetikzlibrary{positioning,arrows.meta,shapes.geometric,fit}

\usepackage{tcolorbox}

\newcommand{\Fields}{\mathcal{F}}
\newcommand{\Tokens}{\mathcal{T}}
\newcommand{\Target}{\mathcal{V}}

\definecolor{GMcolor}{HTML}{2ca02c}
\definecolor{BSL}{HTML}{1f77b4}
\definecolor{BFTS}{HTML}{F5C518}
\definecolor{TKF}{HTML}{ff7f0e}
\definecolor{SYM}{HTML}{8B5CF6}
\definecolor{ASY}{HTML}{2D9D4E}
\definecolor{TEASER_BFTS}{HTML}{984ea3}
\definecolor{TEASER_GM}{HTML}{e41a1c}

\title{TokenFormer: Unify the Multi-Field and  Sequential Recommendation Worlds}

\author{Yifeng Zhou*, Yuehong Hu*, Zhixiang Feng*, Junwei Pan*, Kaihui Wu, Hanyong Li, \and Shangyu Zhang, Shudong Huang, Zhangbin Zhu, Chengguo Yin, Haijie Gu, Jie Jiang}

\affiliation{%
  \institution{Tencent Inc.}
  \city{Shenzhen}
  \country{China}
}

\email{{joefzhou, patrickhu, lionelfeng, jonaspan}@tencent.com}

\renewcommand{\thanks}[1]{}
\makeatletter
\def\@ACM@additionalaffiliation{}
\makeatother

\acmConference[KDD/WWW-style Draft]{Preprint}{2025}{ }
\acmBooktitle{Preprint}
\acmDOI{}
\acmISBN{}

\begin{document}

\begin{abstract}

Recommender systems have historically developed along two largely independent paradigms: feature interaction models for modeling correlations among multi-field categorical features, and sequential models for capturing user behavior dynamics from historical interaction sequences.
Although recent trends attempt to bridge these paradigms within shared backbones, we empirically reveal that naive unifying these two branches may lead to a failure mode of \emph{Sequential Collapse Propagation} (SCP).
That is, the interaction with those dimensionally ill non-sequence fields leads to the dimensional collapse of the sequence features.
To overcome this challenge, we propose \textsc{TokenFormer}, a unified recommendation architecture with the following innovations. 
First, we introduce a \emph{Bottom-Full-Top-Sliding (BFTS) attention scheme}, which applies full self-attention in the lower layers and shrinking-window sliding attention in the upper layers.
Second, we introduce a \emph{Non-Linear Interaction Representation (NLIR)} that applies one-sided non-linear multiplicative transformations to the hidden states. 
Extensive experiments on public benchmarks and Tencent's advertising platform demonstrate state-of-the-art performance, while detailed analysis confirm that \textsc{TokenFormer} significantly improves dimensional robustness and representation discriminability under unified modeling.

\end{abstract}

\maketitle

\let\thefootnote\relax
\footnotetext{* Authors contributed equally to this research.}

\section{Introduction}
\label{sec:intro}

Recommender systems are a foundational infrastructure of the modern digital economy, powering online advertising, feed ranking, e-commerce, and short-video platforms. Their practical importance is evident in both monetization scale and user engagement intensity. According to the IAB/PwC Internet Advertising Revenue Report, U.S. internet advertising revenue reached \$258.6 billion in 2024, up 14.9\% year over year~\cite{iab2024adrev}. At the same time, social and short-video platforms continue to grow rapidly: DataReportal reports 5.24 billion global social media user identities in early 2025~\cite{datareportal2025social}, while TikTok alone reached at least 1.59 billion users via its advertising tools and its Android users spent almost 35 hours on the app in November 2024~\cite{datareportal2025tiktok,datareportal2025topsocial}. Behind this scale lies a core modeling problem: given highly sparse user, item, and context signals together with users' evolving behavior histories, how can we build recommendation models that are both expressive and scalable?

\begin{figure}[t]
    \centering
    \includegraphics[width=\linewidth]{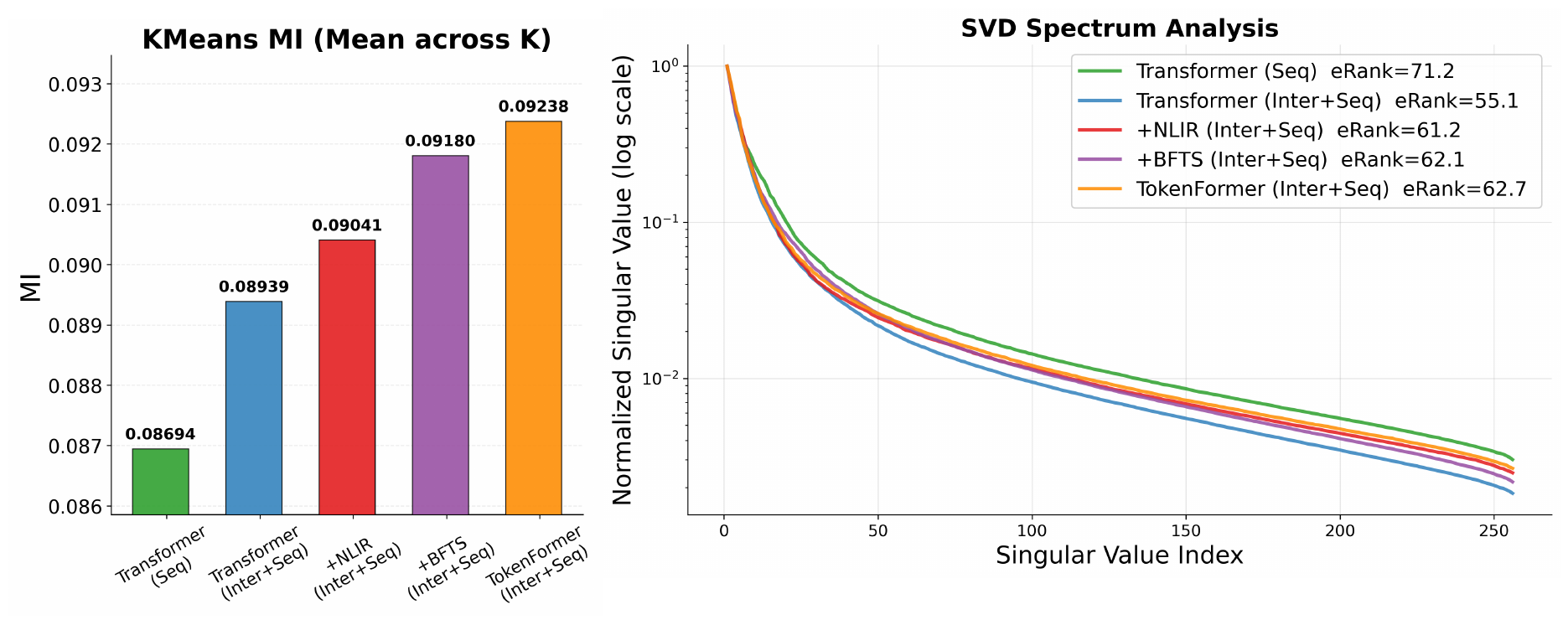}
    \caption{
    The central tradeoff in unified recommendation modeling. 
    Left: representation discriminability measured by mutual information (MI). 
    Right: dimensional robustness measured by the singular value spectrum and effective rank. Compared with a sequence-only Transformer (\textcolor{GMcolor}{green}), naive unification of sequential and non-sequential features with a vanilla Transformer (\textcolor{BSL}{blue}) improves discriminability, but also induces much steeper spectral decay and lower effective rank in sequential representations, revealing \emph{sequential collapse propagation}. 
    The proposed \textcolor{TEASER_BFTS}{BFTS} and \textcolor{TEASER_GM}{NLIR} mitigate this collapse, restore dimensional robustness, and further improve discriminability.
  }
\vspace{-4mm}
  \label{fig:mi_rank}
\end{figure}

A long-standing answer has evolved along two largely separate research branches. 
The first branch focuses on non-sequential \emph{multi-field feature interaction} for sparse tabular recommendation. 
Starting from collaborative filtering and matrix factorization~\cite{koren2009mf}, the literature progressed to explicit interaction models~\cite{rendle2010fm,juan2016ffm,pan2018fwfm,sun2021fm2}, then to deep CTR architectures ~\cite{cheng2016wide,qu2016pnn,guo2017deepfm,he2017nfm,lian2018xdeepfm,wang2021dcnv2,huang2019fibinet}. 
More recently, this line has moved toward dense-scaling interaction backbones~\cite{pan2024ads,zhang2024wukong,zhang2022dhen,zhu2025rankmixer}. 
The second branch focuses on \emph{user interest modeling} from behavior sequences. 
This line spans self-attention sequential recommenders~\cite{kang2018sasrec}, target-aware interest models~\cite{zhou2018din,zhou2019dien,feng2019dsin,zhou2024tin}, and recent long-sequence systems~\cite{pi2020sim,chang2023twin,si2024twinv2,chai2025longer,hu2025tencentlong,hou2024lcn}. 
While both branches are central to industrial recommendation, they have traditionally been developed with different operators, different inductive biases, and different scaling strategies.

However, modern industrial recommender systems are rarely confined to a single modeling paradigm. 
Real-world applications demand holistic reasoning over both static heterogeneous multi-field features and user sequential behavior trajectories.
Despite the clear imperative for unification, bridging these branches remains an open challenge. 
Conventional strategy is to combine them through heterogeneous subnetworks, experts, or late-fusion pipelines~\cite{pan2024ads,zhang2025famousrec, zhang2022dhen}.
Although recent pioneers like InterFormer, OneTrans, HyFormer, and Kunlun~\cite{zeng2025interformer,zhang2025onetr,huang2026hyformer,hou2026kunlun} have moved toward unification, they often still preserve an internal separation through hybrid stacks or alternating components.
Consequently, there is still no fully integrated architecture capable of natively modeling field-field, sequence-sequence, and sequence-field interactions within a single, consistent computational manifold.

A key obstacle to such unification, in our view, is a previously underexplored phenomenon that we term \emph{Sequential Collapse Propagation} (SCP). 
In industrial settings, many non-sequential features have low information abundance due to various reasons, such as low-cardinality or frequency skewness, making their embeddings prone to occupying a low-dimensional subspace~\cite{guo2024embeddingcollapse}. 
In conventional decoupled models, this collapse is safely confined to the non-sequential side. 
However, under a unified model, these collapse-prone static tokens directly interact with sequential behavior tokens through shared operators. 
Fig.\ref{fig:mi_rank} reveals a critical tension in such unification. 
A \textcolor{GMcolor}{sequence-only} Transformer preserves a comparatively high-dimensional sequential representation space, but suffers from weaker discriminability because it does not leverage non-sequential signals. 
A naive \textcolor{BSL}{joint modeling} Transformer over both sequence and non-sequence token streams substantially improves mutual information, indicating that static features provide valuable predictive cues, yet it also exhibits a markedly steeper spectral decay, showing that the sequential representations have become significantly more collapsed. 



To this end, we propose \textsc{TokenFormer} to handle these challenges with three pivotal designs.
First, it unifies static fields, behavior tokens, and target attributes into a monolithic stream, enabling all dependencies to be learned by unified blocks. 
Second, it introduces a \emph{bottom-full-top-sliding} (BFTS) attention schedule, featuring Sliding Window Attention (SWA) with shrinking sliding windows in higher layers.
Third, it employs a \emph{Non-Linear Interaction Representation} (NLIR) gated mechanism on the model representation layers. 
As shown in Figure \ref{fig:mi_rank}, the combination of \textcolor{TEASER_BFTS}{BFTS} and \textcolor{TEASER_GM}{NLIR} \emph{effectively preserves and recovers the intrinsic dimensionality} of the representations, thereby improving discriminability and mitigating the degradation of sequential modeling caused by dimensionally ill static features. We provide detailed analyses in Sec.~\ref{sec:experiments} to further validate these findings.

Extensive experiments on public datasets and Tencent's large-scale online advertising platform demonstrate the effectiveness of \textsc{TokenFormer}. 
Beyond overall offline and online improvements, we perform comprehensive analyses of attention patterns, dimensional robustness and representation discriminability, showing that the proposed architecture not only improves recommendation accuracy but also substantially mitigates the dimensional collapse that arises when sequence and non-sequence tokens are unified. 

The main contributions of this work are summarized as follows:
\begin{itemize}
    \item We identify \emph{Sequential Collapse Propagation} as a central challenge in unified recommendation, and provide empirical evidence showing that collapse-prone non-sequential features can induce dimensional collapse in sequential representations under shared backbones.
    \item We propose \textsc{TokenFormer}, a homogeneous decoder-only architecture that unifies static feature fields and behavior sequences into a single token stream for recommendation.
    \item We introduce a bottom-full-top-sliding interaction hierarchy and a non-linear interaction representation mechanism to jointly improve temporal modeling efficiency, interaction expressiveness, and dimensional robustness.
    \item We validate \textsc{TokenFormer} through extensive offline experiments and online deployment in Tencent Ads, together with detailed analyses that explain why the proposed design better resists collapse than existing alternatives.
\end{itemize}

\section{Related Work}

\subsection{Feature Interaction and Sequential Modeling.} 
Feature interaction in multi-field data has evolved from linear methods and Factorization Machines \cite{richardson2007predicting, cheng2016wide, rendle2010fm, juan2016ffm, pan2018fwfm} to high-order neural crossings like DCN and xDeepFM \cite{wang2017deep, lian2018xdeepfm}. Modern industrial backbones, including Wukong, DHEN, and RankMixer \cite{zhang2024wukong, zhang2022dhen, zhu2025rankmixer}, further scale these interactions for large-scale deployment. 
Recent studies on representation collapse and expressiveness \cite{guo2024embeddingcollapse, pan2024ads} suggest that explicit feature crossing, much like attention-based field models \cite{song2019autoint}, fundamentally follows a pipeline of \emph{projection}, \emph{weighting}, and \emph{multiplicative interaction}.

Parallelly, sequential recommendation has transitioned from recurrent/convolutional units \cite{hidasi2015gru4rec, tang2018caser} to self-attention backbones \cite{kang2018sasrec, sun2019bert4rec} and target-aware architectures \cite{zhou2018din, zhou2019dien, feng2019dsin, pi2020sim, chen2019behavior}. As industrial logs expand, a dedicated line of work scales sequence modeling to lifelong histories via retrieval-based or two-stage systems \cite{pi2020sim, chang2023twin, si2024twinv2, chai2025longer, hu2025tencentlong}. To maintain computational tractability at these scales, sliding-window attention (SWA) \cite{beltagy2020longformer, liu2021swin, Mistral_7B} and localized transducers like HSTU-Ultra \cite{ding2026bending} have emerged as critical primitives for balancing receptive field with efficiency. 

A discernible trend across both paradigms is the increasing adoption of gating mechanisms. Modern architectures, notably HSTU \cite{zhai2024hstu} and its derivatives, demonstrate that augmenting attention with multiplicative modulation significantly bolsters stability and capacity. Under this lens, sequential models converge on a recurring computational motif—computing relevance, aggregating context, and performing gated interactions—revealing that sequential attention is structurally isomorphic to explicit feature interaction, where the ``position'' in sequences plays a role fundamentally analogous to the ``field'' in multi-field data.

\subsection{Representation Collapse in Recommendation}
 
Representation collapse refers to the tendency of learned embeddings to occupy a low-dimensional subspace, losing discriminability across items or users. In the context of recommendation, this problem is especially acute for low-cardinality or weak-information features such as demographic buckets or coarse context fields. Recent work has systematically demonstrated that embedding collapse becomes a critical barrier when scaling up recommendation models~\cite{guo2024embeddingcollapse}, and that collapse and feature entanglement jointly degrade recommendation quality in large-scale industrial systems~\cite{pan2024ads}. From a model design perspective, the Feature Generation paradigm shows that nonlinear interaction operators can explicitly preserve representation rank during feature crossing~\cite{yin2025feature}, providing a theoretical basis for using multiplicative gating to mitigate collapse.
 

\subsection{Toward Unified Modeling of Fields and Sequences}

Despite sharing computational primitives, feature interaction and sequential modeling have largely evolved as separate trajectories. Classical hybrid systems, such as DIN \cite{zhou2018din}, DIEN \cite{zhou2019dien}, DSIN \cite{feng2019dsin}, BST \cite{chen2019behavior}, and SIM \cite{pi2020sim}, typically rely on heterogeneous pipelines that late-fuse cross-feature modules with sequence encoders. While effective, these designs lack a homogeneous architecture capable of jointly modeling field-field, sequence-sequence, and sequence-field dependencies \cite{gui2023HiFormer}.

Recent literature has sought to bridge this gap through diverse architectural innovations. InterFormer \cite{zeng2025interformer} introduces a components interleaving design to facilitate bidirectional information flow, while OneTrans \cite{zhang2025onetr} unifies token types via mixed parameterization. Similarly, HyFormer \cite{huang2026hyformer} employs query decoding to manage interactions, and Kunlun \cite{hou2026kunlun} adopts a per-layer dual-block strategy. Despite these advancements, a fundamental structural bifurcation between token types often persists.

In contrast, \textsc{TokenFormer} originates from a \emph{unified formulation}: we demonstrate that both explicit feature interaction and sequential attention can be subsumed under a single computational template of projection, relevance weighting, and context mixing. 
By treating all fields and behaviors as a singular entity stream within a shared gated-attention operator, \textsc{TokenFormer} provides a more uniform and scalable backbone than prior multi-path architectures.

\section{Problem Setup and Preliminaries}
\label{sec:setup}


We consider recommendation inputs composed of three groups of entities: non-sequential field features \(\Fields\), sequential behavior tokens \(\Tokens\), and target features \(\Target\). Existing recommendation models typically operate on only part of this entity space and focus on specific interaction patterns. This perspective serves as the conceptual bridge to the unified formulation in Sec.~4.1.

\noindent \textbf{Feature interaction models.}
Classical feature interaction models mainly consume \((\Fields,\Target)\). Each feature is first mapped from its vanilla ID to an embedding, and an interaction function is then applied to model correlations among non-sequential fields and target-side features:
\begin{equation}
\hat{y}=\phi\!\left(g_{\mathrm{FI}}(\Fields,\Target)\right),
\end{equation}
where \(g_{\mathrm{FI}}(\cdot)\) denotes the feature interaction function and \(\phi(\cdot)\) denotes the prediction head. Depending on the model, \(g_{\mathrm{FI}}\) may be instantiated as pairwise product, bilinear interaction, or explicit high-order crossing, as in FM, FFM, FwFM, FmFM, CrossNet, and related variants~\cite{rendle2010fm,juan2016ffm,pan2018fwfm,sun2021fm2,wang2021dcnv2,guo2017deepfm,huang2019fibinet}. 

\noindent \textbf{Sequential recommendation models.}
Sequential models mainly consume \(\Tokens\), or \((\Tokens,\Target)\) in the target-aware setting. Each behavior token is first represented by aggregating its constituent features, and the model then captures dependencies through attention-based operators. In particular,
\begin{equation}
\mathbf{H}=g_{\mathrm{self}}(\Tokens),
\end{equation}
where \(g_{\mathrm{self}}(\cdot)\) denotes self-attention over historical behaviors and captures \(\Tokens \leftrightarrow \Tokens\) interactions, as in SASRec, BERT4Rec, BST, and related Transformer-style sequential recommenders~\cite{kang2018sasrec,sun2019bert4rec,chen2019behavior,zhai2024hstu}:
\begin{equation}
\mathbf{u}=g_{\mathrm{target}}(\Tokens,\Target),
\end{equation}
where \(g_{\mathrm{target}}(\cdot)\) denotes target attention and captures \(\Tokens \leftrightarrow \Target\) interactions, as in DIN, DIEN, DSIN, SIM, TWIN, and TIN~\cite{zhou2018din,zhou2019dien,feng2019dsin,pi2020sim,chang2023twin,zhou2024tin}.

\noindent \textbf{Cross-feature sequential modeling.}
A smaller but important line of work additionally introduces interactions between sequential and non-sequential entities. Such models jointly consume \((\Fields,\Tokens)\) and use a cross-feature interaction function
\begin{equation}
\mathbf{z}=g_{\mathrm{cross}}(\Fields,\Tokens),
\end{equation}
to capture \(\Fields \leftrightarrow \Tokens\) dependencies, for example through cross-attention~\cite{zheng2022autoattention,zeng2025interformer}, heterogeneous fusion, or unified hybrid interaction blocks~\cite{zhang2025onetr,huang2026hyformer,hou2026kunlun}.

\noindent \textbf{From partial interactions to unified interaction.}
From this viewpoint, existing paradigms cover only subsets of the full entity space: feature interaction models focus on \((\Fields,\Target)\), self-attention models focus on \(\Tokens\), target-attention models focus on \((\Tokens,\Target)\), and cross-feature sequential models focus on \((\Fields,\Tokens)\). This suggests a more general formulation: recommendation can be cast as learning interactions over the unified entity set \(\Fields \cup \Tokens \cup \Target\). Our model is built on this principle.

\section{TokenFormer}
\label{sec:methodology}

\subsection{Unified Token Stream}

\begin{figure*}[t]
    \centering
    \includegraphics[width=0.96\linewidth]{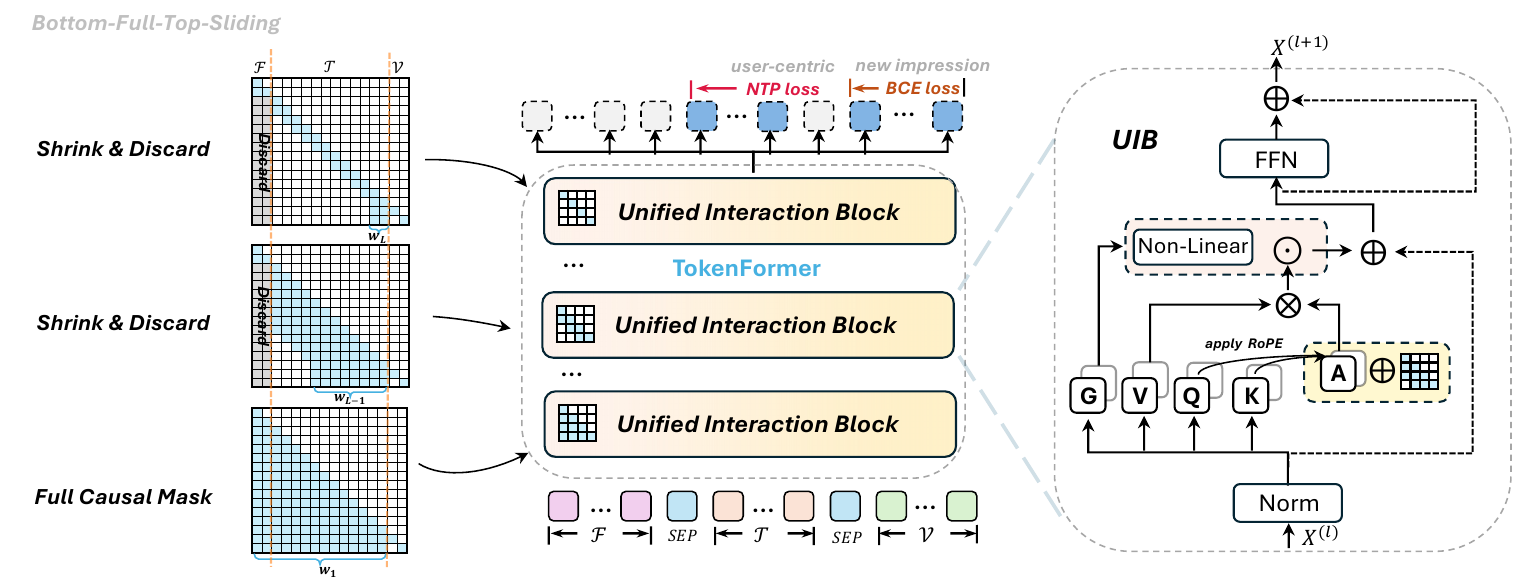}
    \caption{Overview of \textsc{TokenFormer}. 
    TokenFormer represents multi-field features $\Fields$, sequential behavior tokens $\Tokens$, and target features $\Target$ as a unified token stream, which is processed by stacked Unified Interaction Blocks (UIBs). 
    Each UIB combines the proposed \emph{Bottom-Full-Top-Sliding (BFTS)} attention design, which applies full causal attention in shallow layers and shrinking SWA in deeper layers, with the \emph{Non-Linear Interaction Representation (NLIR)} for multiplicative feature interaction. 
    }
    \label{fig:tokenformer}
\end{figure*}

The architecture of \textsc{TokenFormer} is illustrated in Figure~\ref{fig:tokenformer}. We define a unified entity set
\begin{equation}
    \mathcal{E} = \Fields \cup \Tokens \cup \Target.
    \label{eq:entities}
\end{equation}
A unified backbone should jointly model both intra-group and inter-group interactions over \(\mathcal{E}\). These interactions include:
\begin{itemize}
    \item \textbf{\(\Fields \leftrightarrow \Fields\):} correlations among non-sequential user, item, and context features, which are the core object of feature interaction models.
    \item \textbf{\(\Tokens \leftrightarrow \Tokens\):} correlations among historical behaviors, corresponding to self-attention in sequential recommendation.
    \item \textbf{\(\Target \leftrightarrow \Target\):} correlations among target-side features for constructing an expressive target representation.
    \item \textbf{\(\Tokens \leftrightarrow \Target\):} behavior--target relevance, corresponding to target-attention in target-aware recommendation.
    \item \textbf{\(\Fields \leftrightarrow \Tokens\):} dependencies between user behaviors and non-sequential user/context features, typically handled by cross-feature sequential modeling.
    \item \textbf{\(\Fields \leftrightarrow \Target\):} user--target and context--target correlations, which are central to collaborative filtering and feature interaction models.
\end{itemize}
Therefore, unified recommendation requires a single architecture that can simultaneously model all six interaction types, rather than separating feature interaction and sequential modeling into heterogeneous components.

\subsubsection{Unified Entity Stream.}
\textsc{TokenFormer} represents all inputs features as a flattened stream of unified tokens. 
Let $M$ be the number of feature fields, $T$ the historical sequence length, and $K$ the number of target items. The total input sequence length $S_L$ is given by:
\begin{equation}
    S_L =
    \begin{cases}
        M + T + K + N_{\text{sep}} & \text{without actions}, \\
        M + 2T +K + N_{\text{sep}} & \text{with actions},
    \end{cases}
    \label{eq:stream-len}
\end{equation}
where $N_{\text{sep}}$ denotes the number of special delimiter tokens.

Unlike conventional models that employ heterogeneous modules for different data types, we construct a unified input $\mathbf{X}^{(0)} \in \mathbb{R}^{S_L \times d}$ for the initial layer by concatenating all entity embeddings. 
Taking the action-aware scenario as an example, the sequence $\mathbf{X}^{(0)}$ is formulated as:
\begin{equation}
    \mathbf{X}^{(0)} = \Big[ 
    \underbrace{\mathbf{x}^{\mathcal{F}}_{1}, \dots, \mathbf{x}^{\mathcal{F}}_{M}}_{\text{non-seq tokens}}, \mathbf{e}_{\text{sep}},
    \underbrace{\mathbf{x}^{\mathcal{T}}_{s_1}, \mathbf{x}^{\mathcal{T}}_{a_1}, \dots, \mathbf{x}^{\mathcal{T}}_{s_T}, \mathbf{x}^{\mathcal{T}}_{a_T}}_{\text{seq tokens}}, \mathbf{e}_{\text{sep}},
    \underbrace{\mathbf{x}^{\mathcal{V}}_{c_1}, \dots, \mathbf{x}^{\mathcal{V}}_{c_K}}_{\text{target tokens}}
    \Big]^\top,
    \label{eq:unified-x0}
\end{equation}
Notably, we dispense with explicit type embeddings, instead, employ a unified Rotary Positional Embedding (RoPE) across the entire stream to inject relative positional information. 
This allows the model to capture dependencies within a unified geometric space, as further detailed in Sec.~\ref{sec:hybrid-masking}. 
To distinguish between different segments, we insert a special delimiter token $\langle \texttt{sep} \rangle$ (with embedding $\mathbf{e}_{\text{sep}}$) as the segment boundary.

\noindent \textbf{Unified Positional Assignment.} 
To align sequential and non-sequential multi-field tokens within the RoPE-enhanced attention, we propose a type-aware indexing scheme. 
We denote $S_L$ as the maximum sequence length. Static fields are mapped to a shared prefix, while behavioral tokens and the target follow a relative chronological order:
\begin{equation}
p_i = 
\begin{cases} 
0 & \text{if} \; x_i \in \Fields, \\
{pos}(x_i) & \text{if} \;  x_i \in {\Tokens}, \\
S_L + 1 & \text{if} \; x_i \in \Target,
\end{cases}
\label{eq:type-aware-pos}
\end{equation}
where $pos(\cdot)$ extracts the chronological temporal index of the behavioral token.
This design ensures to preserves the invariant properties of static features while maintaining the sensitive temporal evolution of user behaviors.

\subsection{Unified Interaction Block}
\label{sec:hmi-module}

\textsc{TokenFormer} adopts a homogeneous decoder-only backbone with $L$ stacked Unified Interaction Blocks (UIBs). 
Each block is a variant of the vanilla attention block, specifically augmented by two key innovations: (i) a Bottom-Full-Top-Sliding (BFTS) attention mechanism that modifies the attention mask in a layer-dependent manner, and (ii) a nonlinear interacted representation (NLIR) module that improves the attention output through multiplicative modulation. The core transformation at layer $l$ is summarized as:
\begin{align}
    \tilde{\mathbf{A}}^{(l)}
    &= \mathrm{Attn}(\mathbf{X}^{(l)}, M^{(l)}_{BFTS}), \\
    \mathbf{X}^{(l+1)}
    &= \mathrm{FFN}(\mathrm{NLIR}(\tilde{\mathbf{A}}^{(l)}, \mathbf{X}^{(l)})).
    \label{eq:uib-core}
\end{align}
Here, $\mathrm{Attn}(\cdot)$ denotes attention with the proposed BFTS mask schedule ($M^{(l)}_{BFTS}$), and $\mathrm{NLIR}(\cdot)$ denotes the nonlinear interaction applied to the attention output. The detailed formulations of these two components are given in Sec.~\ref{sec:bfts-mask} and Sec.~\ref{sec:gated-mechanism}, respectively.

By stacking such blocks, \textsc{TokenFormer} \emph{progressively integrates global contextual information in shallow layers and emphasizes more localized temporal structure in deeper layers}, while simultaneously enhancing the expressiveness of the attention output through nonlinear interaction.

\subsection{BFTS Attention Mechanism}
\label{sec:bfts-mask}
We next describe the attention mechanism used in the Unified Interaction Block. Given the normalized input $\mathbf{X}^{(l)}$ at layer $l$, we first project it into $\mathbf{Q}$, $\mathbf{K}$, and $\mathbf{V}$, and compute attention in the standard causal form. 
In \textsc{TokenFormer}, we adopt Rotary Position Embedding (RoPE) to encode relative order information, yielding the attention output:
\begin{equation}
    \tilde{\mathbf{A}}^{(l)} = \mathrm{Softmax}\!\left(
    \frac{\mathcal{R}(\mathbf{Q}, \Theta)\,\mathcal{R}(\mathbf{K}, \Theta)^\top}{\sqrt{d_k}}
    + \mathcal{M}^{(l)}
    \right)\mathbf{V},
    \label{eq:ugi-attn}
\end{equation}
where $\mathcal{R}(\cdot,\Theta)$ denotes the RoPE transformation, and $\mathcal{M}^{(l)}$ is the visibility mask at layer $l$. 

\noindent \textbf{Sliding Window Attention.}
A natural way to reduce the cost of dense attention on long behavioral sequences is to restrict each token to a local receptive field. Concretely, in sliding-window attention (SWA), token $i$ is allowed to attend only to the most recent $w$ valid predecessors. This is implemented by the following mask:
\begin{equation}
    \mathcal{M}^{(l)}_{i,j} =
    \begin{cases}
    0, & \text{if } j \le i \text{ and } i-j < w,\\
    -\infty, & \text{otherwise}.
    \end{cases}
    \label{eq:swa-mask}
\end{equation}
Compared with full causal attention, SWA reduces the effective attention range from the entire prefix to a local window, thereby focusing the model on nearby temporal dependencies.

SWA is particularly attractive in unified recommendation modeling for two reasons. First, it reduces the computational and memory cost of attention from dense global interactions to sparse local interactions, which is important when the unified token sequence contains long user histories. Second, many fine-grained behavioral dependencies in recommendation are inherently local, such as short-term interest continuation and near-neighbor behavior co-occurrence. By constraining attention to a sliding window, the model can emphasize such local temporal patterns while suppressing interference from distant and potentially noisy behaviors.

\noindent \textbf{From Uniform SWA to Bottom-Full-Top-Sliding.}
Despite these advantages, applying SWA uniformly to all layers is suboptimal for a unified architecture. In \textsc{TokenFormer}, the input sequence contains not only sequential behavior tokens but also heterogeneous static feature tokens. Early layers therefore need sufficiently broad receptive fields to establish global cross-domain interactions among these heterogeneous tokens. If all layers are restricted to local windows from the beginning, the model may prematurely lose the ability to propagate global contextual information across the unified sequence.

To balance global interaction and local refinement, we propose the \emph{Bottom-Full-Top-Sliding} (BFTS) mechanism. 
The key idea is simple: shallow layers use full causal attention to build a comprehensive cross interaction over the unified token stream, while deeper layers switch to shrinking sliding-window attention to refine local temporal structure on top of that foundation. 
This mechanism aligns the attention range with representation depth: broad interaction is emphasized in lower layers, while localized refinement is emphasized in deeper layers.

Formally, let the $L$-layer backbone consist of $l_{\mathrm{f}}$ full-attention layers and $l_{\mathrm{s}}$ sliding-window-attention layers, where $L=l_{\mathrm{f}}+l_{\mathrm{s}}$. The entire transformation can be written as
\begin{equation}
    \mathbf{X}^{(L)}
    =
    \underbrace{\left(\mathcal{F}_{\mathrm{SWA}}^{(w_l)} \circ \dots \circ \mathcal{F}_{\mathrm{SWA}}^{(w_0)}\right)}_{l_{\mathrm{s}} \text{ layers}}
    \circ
    \underbrace{\left(\mathcal{F}_{\mathrm{Full}}^{(\infty)} \circ \dots \circ \mathcal{F}_{\mathrm{Full}}^{(\infty)}\right)}_{l_{\mathrm{f}} \text{ layers}}
    \bigl(\mathbf{X}^{(0)}\bigr),
    \label{eq:bfts-composition}
\end{equation}
where $\mathcal{F}$ denotes the layer transformation function. 
To implement the \textit{shrinking window} strategy, the attention spans $\{w_k\}_{k=1}^{l_{\mathrm{s}}}$ are ordered as $w_{l_{\mathrm{s}}} < w_{l_{\mathrm{s}}-1} < \dots < w_1$.
Accordingly, the layer-dependent visibility mask is defined as
\begin{equation}
    \mathcal{M}^{(l)}_{i,j} =
    \begin{cases}
    0, & \text{if } j \le i \text{ and } i-j < \omega(l),\\
    -\infty, & \text{otherwise},
    \end{cases}
    \qquad
    \omega(l)=
    \begin{cases}
    \infty, & l \le l_{\mathrm{f}},\\
    w_l, & l > l_{\mathrm{f}}.
    \end{cases}
    \label{eq:bfts-mask}
\end{equation}
This progressive reduction in window size forces the model to distill broad global dependencies into increasingly granular and localized representations.
\textsc{TokenFormer} preserves global context modeling where it is most needed and introduces local sparsification only after the model has established sufficiently rich cross-token interactions.

\noindent \textbf{Non-Sequence Token Discarding.}
\label{sec:token-discard}
In our unified stream, the static field tokens $X^{\mathcal{F}}$ primarily serve as global contextual priors. 
Once their information is sufficiently integrated into the sequence representations during the initial layers, retaining them becomes redundant for further refinement. 
To enforce a more effective representation paradigm, we implement a layer-wise token discarding strategy. 
Specifically, after layer $l_{\text{f}}$, the model completely ceases to attend to the first $M$ non-sequence tokens:
\begin{equation}
\mathcal{M}^{(l)}_{i,j} = -\infty, \quad \forall\, l \ge l_{\text{f}}, \quad i \in [M, S_L-1], j \in [0, M-1].
\label{eq:discard}
\end{equation}
This strategy ensures that deeper layers allocate their full expressive capacity and attention bandwidth solely to behavioral evolution and target-aware reasoning, further forcing the cross-feature interactions (between static features and sequential behaviors) to be thoroughly completed within the early layers.

\subsection{Non-Linear Interacted Representation}
\label{sec:gated-mechanism}
\textsc{TokenFormer} incorporates a unified nonlinear interaction paradigm~\cite{huang2019fibinet, PEPNet, yin2025feature} designed to enhance \textit{representational discriminability} and recover \textit{dimensional robustness}. 
Unlike conventional gating mechanisms that treat the modulation branch as a passive coefficient, we interpret it as a learned nonlinear transformation that interacts multiplicatively with the primary feature stream. 
This design choice aims to strengthen feature expressiveness while simultaneously mitigating representation collapse.

\noindent \textbf{Multiplicative Interaction in Attention.} 
To modulate the dependency between dense behavioral signals and sparse static fields, we apply an element-wise interaction operation to the attention output $\mathbf{A}^{(l)}$. 
Concretely, we first compute a gate projection from the layer input:
\begin{equation}
    \mathbf{G}^{(l)} = \mathbf{X}^{(l)} \mathbf{W}_{g}^{(l)}, 
    \qquad 
    \mathbf{W}_{g}^{(l)} \in \mathbb{R}^{d \times d},
    \label{eq:gate-proj}
\end{equation}
and then use it to modulate the attention output:
\begin{equation}
    \tilde{\mathbf{I}}^{(l)} = \sigma(\mathbf{G}^{(l)}) \odot \mathbf{A}^{(l)},
    \label{eq:ugi-gate}
\end{equation}
where $\sigma(\cdot)$ is the sigmoid function. 
This operation introduces high-order non-linearity into the token mixing process, thereby strengthening feature interactions and enhancing the discriminability and diversity of the latent representations. 
By modulating the attention output through multiplicative gating, the model effectively preserves the rank richness of the feature space, producing a more expressive interacted representation for the subsequent stage.
The corresponding post-attention residual state is:
\begin{equation}
    \mathbf{I}^{(l)} = \mathbf{X}^{(l)} + \tilde{\mathbf{I}}^{(l)}.
    \label{eq:attn-residual}
\end{equation}

\subsection{SwiGLU Feed-Forward Network.}
Following the gated attention operation, \textsc{TokenFormer} employs a SwiGLU-based feed-forward network. Given the interacted representation $\mathbf{I}^{(l)}$, we first apply RMSNorm and then compute
\begin{equation}
\label{eq:swiglu-rmsnorm}
\begin{aligned}
    &\tilde{\mathbf{I}}^{(l)} = \mathrm{RMSNorm}(\mathbf{I}^{(l)}), \\
    &\mathbf{H}^{(l)} = \Big(
    \mathrm{Swish}(\tilde{\mathbf{I}}^{(l)} \mathbf{W}_1)
    \odot
    (\tilde{\mathbf{I}}^{(l)} \mathbf{W}_2)
    \Big)\mathbf{W}_3, \\
    &\mathbf{X}^{(l+1)} = \mathbf{I}^{(l)} + \mathbf{H}^{(l)},
\end{aligned}
\end{equation}
where $\mathbf{W}_1$, $\mathbf{W}_2$, and $\mathbf{W}_3$ are learnable weight matrices. This design preserves the standard residual feed-forward update while keeping the feed-forward stage consistent with the multiplicative interaction principle used in the attention branch.

\subsection{Unified Optimization Objectives}
\label{sec:supervision-paradigm}

A distinguishing feature of \textsc{TokenFormer} is that the same unified token architecture can support different recommendation paradigms under a shared supervision framework. 
In this work, we consider two settings that are also used in the experiments: \textit{User-Centric} recommendation and \textit{New Impression Only} recommendation.

\noindent \textbf{User-Centric setting.} The model is trained with dense autoregressive supervision over the unified token sequence. Historical behaviors, static user fields, and candidate-related tokens are jointly modeled, and the supervision signal can be applied throughout the sequence in a next-token style manner. This setting encourages the model to learn a comprehensive user representation from the full interaction context.

\noindent \textbf{New Impression Only setting.} The model focuses on targeted supervision over the newly exposed impression tokens, while treating the preceding user history as contextual input. 
Historical interactions serve as auxiliary contextual priors, providing the necessary background information without contributing to the loss gradients. 
Consequently, the loss is applied only to the candidate items or terminal decision tokens associated with the current impression thereby aligning this setup with common industrial ranking scenarios where the prediction target is confined to the newly arriving candidate set.


Formally, let $\Omega = \{1, \dots, S_L\}$ be the set of indices for the unified stream. 
We define $\mathcal{I}_{\text{loss}} \subset \Omega$ as the subset of indices designated for supervision (e.g., indices of target candidates). 
For each supervised index $t \in \mathcal{I}_{\text{loss}}$, the final output representation $\mathbf{X}^{(L)}_t \in \mathbb{R}^d$ is projected onto an $A$-dimensional action space via a linear head:
\begin{equation}
    \boldsymbol{\ell}_t = \mathbf{W}_L \mathbf{X}^{(L)}_t + \mathbf{b}_L, \quad \boldsymbol{\ell}_t \in \mathbb{R}^{N},
    \label{eq:rank-logits}
\end{equation}
where $\mathbf{W}_L \in \mathbb{R}^{N \times d}$ and $\mathbf{b}_L \in \mathbb{R}^N$ are learnable parameters. 
The logits are then normalized with a softmax function, and the model is trained with the Cross-Entropy (CE) loss:
\begin{equation}
    \mathcal{L}_{\text{CE}} = -\frac{1}{|\mathcal{I}_{\text{loss}}|} \sum_{t \in \mathcal{I}_{\text{loss}}} \log \left( \frac{\exp(\ell_{t, c_t})}{\sum_{a=1}^A \exp(\ell_{t, a})} \right),
    \label{eq:rank-ce}
\end{equation}
where $c_t \in \{1, \dots, N\}$ denotes the ground-truth action label for the token at position $t$.

\section{Experiments}\label{sec:experiments}
We conduct extensive experiments to rigorously evaluate the effectiveness, efficiency, and internal mechanisms of our proposed \textsc{TokenFormer}. 
To provide a clear roadmap for our empirical analysis, we aim to answer the following core Research Questions (RQs) in this section:

\begin{itemize}[leftmargin=*]
    \item \textbf{RQ1 (Overall Performance):} How does \textsc{TokenFormer} perform against state-of-the-art baselines under different modeling paradigms (i.e., User-Centric and New Impression Only)? (Detailed in Sec.~\ref{sec:exp_overall})

    \item \textbf{RQ2 (Representational Discriminability):} Do the NLIR and explicit BFTS constraints substantially enhance the discriminability of the final representations? (Detailed in Sec.~\ref{sec:exp_gated_expressiveness})

    \item \textbf{RQ3 (Dimensional Robustness):} How the integration of NLIR and BFTS mitigates the Sequential Collapse Propagation (SCP) typically induced by unifying high-dimensional sequential behaviors with low-dimensional static features? (Detailed in Sec.~\ref{sec:exp_gated_dimension}).
    
    \item \textbf{RQ4 (Layer-wise Attention Allocation):} How does the BFTS allocate attention across layers, assigning global heterogeneous integration to shallow layers and localized temporal refinement to deep layers? (Detailed in Sec.~\ref{sec:exp_bfts_motivation})
    
    \item \textbf{RQ5 (Efficiency and Effectiveness):} Can BFTS-based SWA effectively reduce computational cost (GFLOPs) while simultaneously boosting predictive accuracy? (Detailed in Sec.~\ref{sec:exp_bfts_dual_benefits}).
    
    \item \textbf{RQ6 (Ablation Study):} What are the individual contributions of each core architectural component within \textsc{TokenFormer}? (Detailed in Sec.~\ref{sec:exp_ablation})
    
    \item \textbf{RQ7 (Industrial Scalability):} Does \textsc{TokenFormer} adhere to neural scaling laws, and how does its representation capacity scale when transitioned from academic datasets to massive industrial environments? (Detailed in Sec.~\ref{sec:scaling_law})
\end{itemize}
\subsection{Experimental Setup}
\label{sec:exp_setting}

\noindent \textbf{Datasets.} 
We evaluate our proposed framework on the publicly available \textit{KuaiRand-27K} dataset, a representative benchmark for sequential recommendation that comprises approximately 27k unique user interaction trajectories. 
To ensure a rigorous assessment of model performance, we partition the dataset into training, validation, and testing sets following a 19k/2k/5k split, respectively. 
To further demonstrate the scalability and generalizability of \textsc{TokenFormer} in production-grade environments, we extend our evaluation to several high-traffic scenarios within the industrial-scale \textit{Tencent Ads platform}. 
Unlike public benchmarks, these industrial datasets encompass billions of interaction logs, characterized by extremely high feature sparsity and dynamic user intent. 

\noindent \textbf{Training Paradigm and Optimization.} 
For parameter optimization, we employ the AdamW optimizer with an initial learning rate of 0.001. 
We adopt a \textit{Next-Token Prediction} (NTP) training paradigm, which enables the model to effectively capture the latent transition dynamics across the entire historical sequence.

\begin{table*}[t]
\centering
\small
\caption{Main results on KuaiRand-27k. $\Delta$ denotes the AUC improvement relative to the Transformer baseline in each category, measured in per mille (\textperthousand).}
\label{tab:main_results}
\renewcommand{\arraystretch}{1.15}
\setlength{\tabcolsep}{8pt} 
\begin{tabular}{llcccc}
\toprule
\textbf{Paradigm} & \textbf{Model} & \textbf{Macro AUC} & \textbf{$\Delta$ (\textperthousand)} & \textbf{Params} & \textbf{GFLOPs}\\
\midrule
\multirow{7}{*}{\shortstack[l]{User-Centric}} 
& Transformer    & 0.85467 & - & 3.41M & 3.7G \\
& HSTU           & 0.85694 & +2.27 & 0.50M & 0.3G \\
& HSTU-Ultra     & 0.85762 & +2.95 & 0.50M & 0.2G \\
\cmidrule(lr){2-6}
& \textbf{TokenFormer-T} & \textbf{0.85967} & \textbf{+5.00} & 0.48M & 0.8G \\
& \textbf{TokenFormer-S} & \textbf{0.86043} & \textbf{+5.76} & 3.77M & 3.5G \\
& \textbf{TokenFormer-M} & \textbf{0.86116} & \textbf{+6.49} & 5.64M & 5.3G \\
& \textbf{TokenFormer-L} & \textbf{0.86282} & \textbf{+8.15} & 10.14M & 9.8G \\
\midrule
\multirow{4}{*}{\shortstack[l]{New Impression Only}} 
& Transformer* & 0.84019 & - & 3.41M & 3.7G \\
& OneTrans       & 0.84663 & +6.44 & 3.91M & 0.1G \\
& HyFormer       & 0.85063 & +10.44 & 2.78M & 0.2G \\
& \textbf{TokenFormer-S}* & \textbf{0.85161} & \textbf{+11.42} & 3.77M & 3.5G \\
\bottomrule
\end{tabular}
\end{table*}
\subsection{Overall Performance Comparison}
\label{sec:exp_overall}

To comprehensively evaluate the effectiveness of \textsc{TokenFormer}, we conduct extensive experiments against a wide range of state-of-the-art baselines. 
Following the data organization principles in recent generative recommendation \cite{onerec}, we categorize these models into two distinct paradigms based on their training objectives and sequence handling:

\noindent \textbf{User-Centric Organization.} 
Models in this setting, including HSTU, HSTU-Ultra, and the vanilla Transformer, are optimized via the standard Next-Token Prediction (NTP) objective across the entire interaction history. This paradigm treats the user's complete behavior sequence as a single coherent training sample. While it strictly preserves chronological dependencies and captures step-by-step auto-regressive transitions, it often suffers from significant computational redundancy as historical patterns are repeatedly processed during training.

\noindent \textbf{New Impression Only Organization.} 
Models in this setting, such as OneTrans, HyFormer, and the target-optimized variants (Transformer* and TokenFormer-S*), are trained exclusively to predict the latest impressed target item. 
In this paradigm, historical behaviors are treated primarily as contextual features to facilitate the interaction with the specific target item, rather than modeling the step-by-step sequential evolution of the entire history. 

Table~\ref{tab:main_results} summarizes the overall performance on the KuaiRand-27K dataset. Based on these empirical results, we draw the following key observations:
\textsc{TokenFormer} significantly outperforms all baselines within the sequence-preserving category. 
Notably, the tiny version outperforms the Transformer baseline by \textbf{5.00‰} in AUC, surpassing the strong baseline HSTU-Ultra by \textbf{2.05‰}.
This superiority underscores the efficacy of our unified gated backbone in capturing heterogeneous feature interactions while maintaining strict sequential dependencies.
A performance gap is observed between the two paradigms. 
Models optimized with NTP loss generally exhibit higher AUC than their target-loss counterparts. 
This suggests that preserving the ordinal consistency of user sequences provides a richer supervisory signal for representation learning. 
Meanwhile, \textsc{TokenFormer} delivers highly competitive performance under the \emph{New Impression Only} setting as well, demonstrating its seamless adaptability across different loss formulations.

\subsection{Impact of NLIR and BFTS on Representation Expressiveness}
\label{sec:exp_gated_expressiveness}

\begin{figure}[t]
    \centering
    \includegraphics[width=0.92\linewidth]{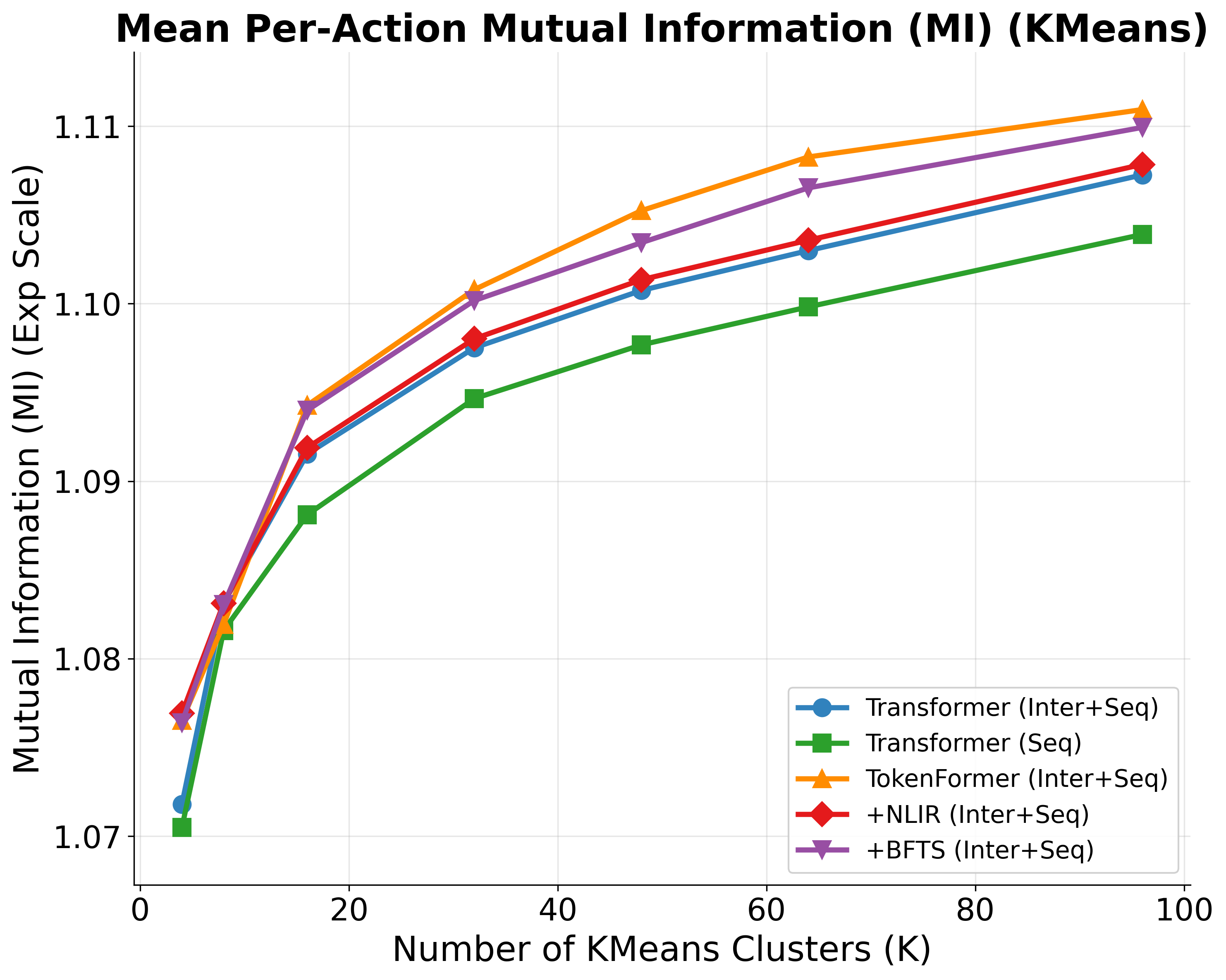}
    \vspace{-2mm}
    \caption{Discriminability analysis across varying cluster numbers $K$. 
    To clearly illustrate the performance gains, we report the scaled MI.
    The proposed BFTS and NLIR can effectively improve discriminability for output embeddings.}
    \vspace{-2mm}
    \label{fig:mi_two}
\end{figure}

To empirically validate the representational advantages of the multiplicative interactions introduced by the Non-Linear Interaction Representation (NLIR), we analyze the Mutual Information (MI) between the post-attention representations and the target labels. 
Intuitively, a higher MI indicates that the learned embeddings retain more discriminative information.
Since the representations are high-dimensional and continuous, we adopt a discretization-based estimator to ensure tractability. 
Specifically, we partition the representation space into $K$ clusters via K-Means and compute the MI between the cluster assignments and the ground-truth labels. 
This metric serves as a direct quantitative metric for the \textit{expressiveness} of the output embeddings.
The results of this discriminability analysis are illustrated in Figure~\ref{fig:mi_two}, where we evaluate the MI across various discretization granularities ($K$). 
While the performance gap remains relatively narrow at smaller values of $K$, it widens significantly as the discretization granularity increases.

The visualization clearly demonstrates the necessity of our proposed components.
Specifically, the integration of the \textcolor{TEASER_BFTS}{BFTS} and the \textcolor{TEASER_GM}{NLIR} yields a consistent and substantial improvement in MI across nearly all values of $K$, confirming that both modules are indispensable for capturing label-predictive features. 
This phenomenon corroborates our theoretical intuition: the explicit non-linear multiplicative interactions introduced by NLIR significantly enrich the representational capacity. 
BFTS provides an optimized interaction pattern to establish a structured receptive field to constrain cross-token interference, while NLIR serves as a critical role for distilling complex signals into high-fidelity representations.
Their complementary interplay \textcolor{TKF}{\textsc{TokenFormer}} significantly bolsters the discriminative power of the final representations.

\begin{figure}[t]
    \centering
    \includegraphics[width=\linewidth]{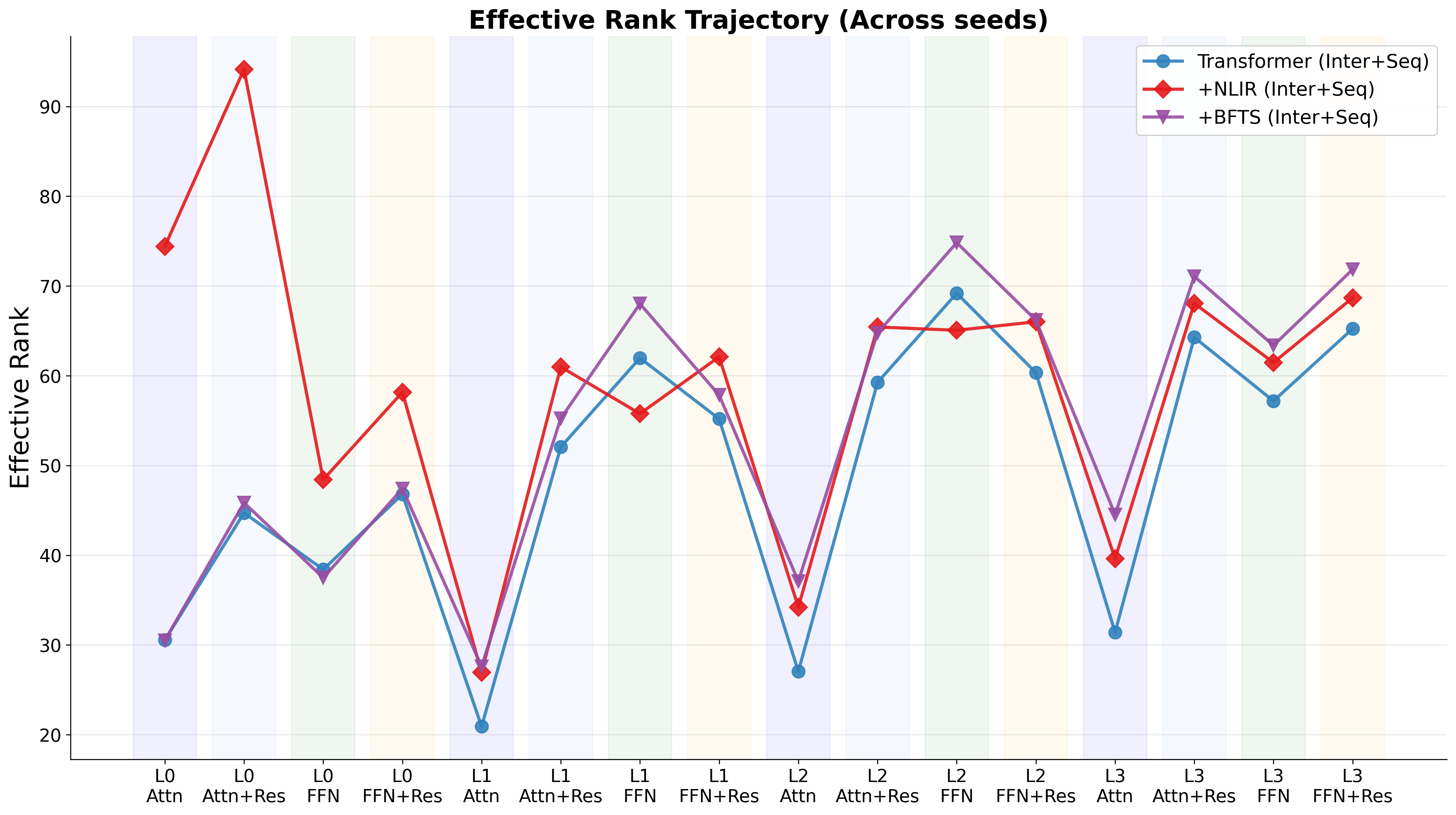}
    \vspace{-4mm}
    \caption{Block-wise effective-rank trajectory of sequential behavioral tokens ($\Tokens$) for the \textcolor{BSL}{Vanilla Transformer}, \textcolor{TEASER_BFTS}{$+$BFTS} and \textcolor{TEASER_GM}{$+$NLIR}. 
    We track the output across each block include: Attention output (Eq.\ref{eq:ugi-attn}), the successive residual additions (Eq.\ref{eq:attn-residual}), and the FFN output (Eq.\ref{eq:swiglu-rmsnorm}).    
    }
    \vspace{-2mm}
    \label{fig:eff_rank_traj}
\end{figure}

Beyond the gains attributable to BFTS and NLIR, Figure 3 also reveals that relying solely on behavioral sequences is insufficient for robust preference modeling: even the vanilla joint modeling (\textcolor{BSL}{Transformer}) yields a substantial gain in representational discriminability over its sequence-only counterpart (\textcolor{GMcolor}{Transformer}), confirming that non-sequential static attributes provide indispensable categorical priors that anchor the evolving user interests.
However, such a naive unification approach cause the \emph{Sequential Collapse Propagation} (SCP), as it fails to preserve the  dimensional robustness for the vulnerable sequential tokens.

\subsection{Impact of NLIR and BFTS on Dimensional Robustness}
\label{sec:exp_gated_dimension}

We hypothesize that the dimensional collapse in unified recommendation is not inevitable. 
Specifically, we conjecture that the Bottom-Full-Top-Sliding (BFTS) architecture and Non-Linear Interaction Representation (NLIR) can separately mitigate this issue: BFTS prevents low-rank noise propagation by varying attention scope across layers, while NLIR enhances dimensional robustness by injecting non-linearity into ill-conditioned static features.

To examine these effects, we conduct a layer-wise spectral analysis on the KuaiRand-27k dataset. 
We compare three model variants: (1) \textcolor{BSL}{Transformer}, a vanilla joint modeling baseline; (2) \textcolor{TEASER_BFTS}{BFTS-only}, which employs global causal attention in the first two layers and shrinking window SWA in the deep layers; (3) \textcolor{TEASER_GM}{NLIR-only}, which conduct the Non-Linear Interaction Representation within the attention block. 
We utilize the effective rank (erank) of hidden representations in each block as a quantitative metric for representational robustness. The detailed computation procedure is provided in Appendix~\ref{app:spectral_diagnostics}.

\begin{figure}[t]
    \centering
    \includegraphics[width=0.95\linewidth]{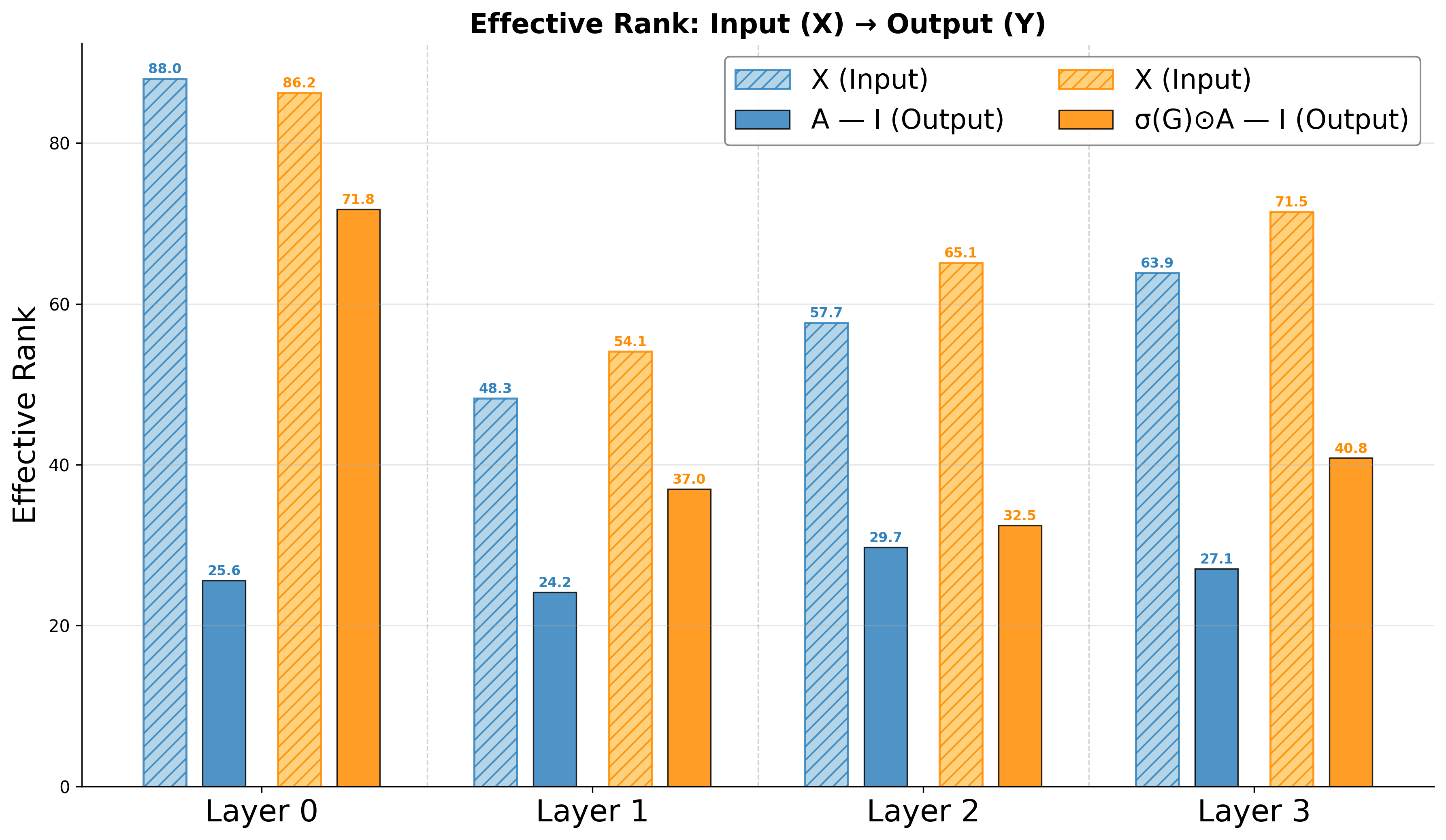}
    \caption{Effective rank comparison of sequential behavioral tokens ($\Tokens$) across layers. 
    To demonstrate the protective effect of the non-linear gating mechanism, we compare the layer-wise outputs given the input $\mathbf{X}^{(l)}$. 
    While the linear attention \textcolor{BSL}{output} (Eq.~\ref{eq:ugi-attn}) suffers from sharply rank degradation, the non-linear gated \textcolor{TKF}{output} of \textsc{TokenFormer} (Eq.~\ref{eq:ugi-gate}) significantly \textbf{mitigates} this collapse in the early stages.}
    \label{fig:effective_rank_xgy_summary2}
\end{figure}

As illustrated in Figure~\ref{fig:eff_rank_traj}, the vanilla \textcolor{BSL}{joint modeling} exhibits a markedly steeper spectral decay, confirming that sequential representations are severely collapsed by the introduction of low-rank static features.
Both {BFTS} and {NLIR} independently contribute to dimension recovery. 
These observations suggest that:
\textcolor{TEASER_BFTS}{BFTS} limits the propagation of collapse by imposing localized attention priors in deeper layers, which prevents low-rank static noise from diluting the high-frequency behavioral signals.
\textcolor{TEASER_GM}{NLIR} restores the representation rank by introducing non-linear multiplicative interactions that promote feature decorrelation. 
This mechanism ensures that the model captures highly expressive representation rather than confining the representation to a few dominant simplistic feature, thereby preserving the expressive granularity.

Beyond the primary architectural components, the spectral trajectories in Figure~\ref{fig:eff_rank_traj} elucidate how the residual connections systematically modulate the representation rank:

\noindent \emph{Attention-Residual Restoration}: The residual connection after the attention operator, in Eq.\ref{eq:attn-residual}: $\mathbf{I}^{(l)} = \mathbf{X}^{(l)} + \tilde{\mathbf{I}}^{(l)}.$, consistently increases the effective rank. 
That is, $\text{erank}(\mathbf{I}^{(l)})$ is systematically higher than $\text{erank}(\tilde{\mathbf{I}}^{(l)})$ across nearly all measured stages. 
This suggests that the attention residual branch does not merely stabilize optimization, but also actively restores dimensional diversity lost during the pure attention transformation.

\noindent \emph{FFN-Residual Regularization}: After the FFN transformation, the subsequent residual connection in Eq.\ref{eq:swiglu-rmsnorm}: $\mathbf{X}^{(l+1)} = \mathbf{I}^{(l)} + \mathbf{H}^{(l)}$, tends to pull the effective rank back to an intermediate range. 
In most cases, $\text{erank}(\mathbf{X}^{(l+1)})$ is located between $\text{erank}(\mathbf{I}^{(l)})$ and $\text{erank}(\mathbf{H}^{(l)})$. 
This pattern suggests the FFN residual branch plays a critical regularizing role, preventing the representation from drifting too far from the rank profile established by the preceding attention stage.

\begin{figure*}[t]
\vspace{-5mm}
    \centering
    \includegraphics[width=0.93\linewidth]{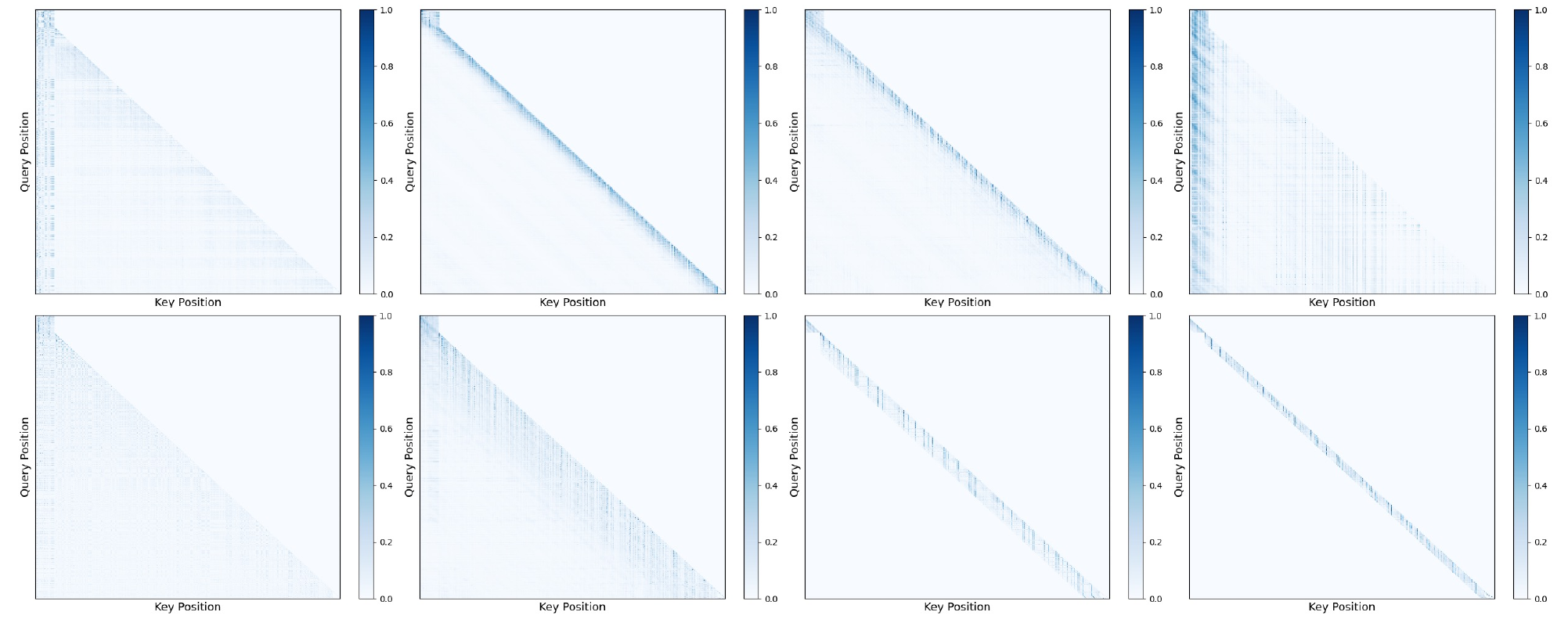}
\vspace{-3mm}
    \caption{Evolution of attention patterns. 
    Top: Vanilla Transformer suffers from redundant revisiting of static fields in last layers; once the initial cross-interactions is completed in shallow layers, such back-attend introduces noise.
    Bottom: \textsc{TokenFormer} decouple this by BFTS. It establishes cross-interactions in shallow layers and switches to shrinking window SWA in deeper layers to refine sequential representations.
    }
    \label{fig:attn_evolution_grid}
\end{figure*}

\begin{figure}[t]
    \centering
    \includegraphics[width=\linewidth]{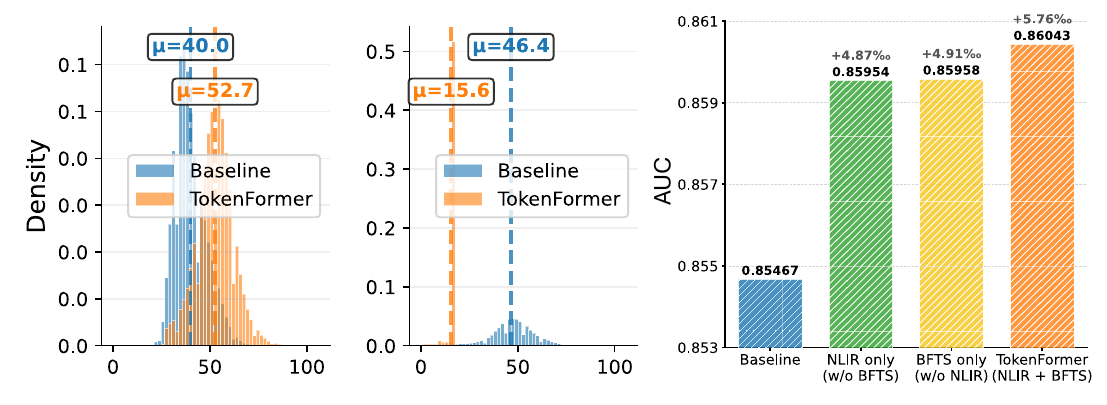}
\vspace{-5mm}
    \caption{Left: Attention receptive field distributions . Histograms left to right correspond to first and last layer, respectively.  
    In shallow layers, \textcolor{TKF}{\textsc{TokenFormer}} exhibits a wider receptive field to facilitate comprehensive cross-integration. 
    In deeper layers, while the \textcolor{BSL}{Transformer} maintains a broad distribution due to redundant back-attention to static tokens.
    Right: AUC for ablation study.
    }
\vspace{-5mm}
    \label{fig:receptive_field_histograms_and_abl}
\end{figure}

We further focus on the sequential behavioral tokens, whose representations are more susceptible to collapse propagation from static fields in the unified architecture. 
As shown in Figure~\ref{fig:effective_rank_xgy_summary2}, although the effective rank decreases gradually with depth in both models, this degradation is substantially mitigated when \textcolor{TKF}{NLIR} is enabled. 
In contrast, the \textcolor{BSL}{vanilla variant} exhibits a sharper drop, suggesting that sequential representations become progressively compressed into a lower-dimensional space.

Taken together, these observations are consistent with our hypothesis. 
They suggest that the non-linear interaction acts as an effective safeguard against dimensional collapse during heterogeneous token interaction. 
More importantly, the results imply that the benefit of gating is not limited to local feature filtering; it also helps preserve representation diversity throughout the network. Meanwhile, the residual branches exhibit their own geometric effect: the attention residual consistently lifts effective rank, while the FFN residual typically re-centers it to an intermediate level. This provides a more refined geometric explanation for the downstream ranking gains of \textsc{TokenFormer}.

\subsection{Layer-wise Attention Analysis in BFTS}
\label{sec:exp_bfts_motivation}

We argue that an effective feature interaction paradigm should allocate distinct attention scopes to different layers.  
Intuitively, the learning process can be viewed as an information refinement pipeline: the network should first perform a comprehensive, global interaction between static (low-dimensional, non-sequential) user profiles in the early stages. Once this global context is established, the deeper layers should exclusively concentrate on localized, temporal fusion without being distracted by static priors.

Specifically, we visualize the attention masks of a {Vanilla Transformer} and our {\textsc{TokenFormer}} across different layers in Figure~\ref{fig:attn_evolution_grid}, and further quantify their corresponding attention receptive field distributions via histograms in Figure~\ref{fig:receptive_field_histograms_and_abl}.

Our visualizations reveal a stark contrast in their interaction behaviors. 
As shown in the attention grids Figure~\ref{fig:attn_evolution_grid} (where the first $M$ tokens denote non-sequential tokens), the {Vanilla Transformer} (top row) attends heavily to these $M$ static tokens in both the initial and final layers, while intermediate layers default to localized band-diagonal interactions.
The final layers exhibit an attention "drift," persistently attending back to distant, non-sequential positions. 
This counter-intuitive behavior is corroborated by the histogram analysis (Figure~\ref{fig:receptive_field_histograms_and_abl}), where the \textcolor{BSL}{Vanilla Transformer} average receptive field paradoxically expands in the final layers (increasing from 40.0 in the intermediate stage to 46.4). 
Conversely, due to the explicit formulation of BFTS, Figure~\ref{fig:attn_evolution_grid}  (bottom row) \textcolor{TKF}{\textsc{TokenFormer}} its shallow layers force significantly richer and broader interactions with the $M$ non-sequential tokens in the shallow layers compared to the \textcolor{BSL}{Vanilla Transformer} (\textcolor{TKF}{52.7} vs. \textcolor{BSL}{40.0}), ensuring a more exhaustive cross-domain feature interaction in shallow layers. 
Furthermore, it exhibits striking \textit{intra-window sparsity} within the narrowed band, dynamically attending only to the most relevant adjacent tokens rather than uniformly weighting the entire local window.
Notably, in deep layers, \textsc{TokenFormer} completely drops the attention to non-sequential positions when applying the SWA.

Consequently, the superior predictive performance (i.e., higher AUC) of \textsc{TokenFormer} over the baseline confirms our initial assumption. 
We argue that once static priors are adequately fused with the behavioral sequence in early stages, repeatedly aggregating them in deep layers becomes redundant, potentially introduces noise, and severely dilutes the model's focus on highly-predictive temporal dynamics. 
By utilizing the band-diagonal SWA pattern to mask out non-sequential interactions, BFTS forcibly contracts the attention scope and discards the structural noise introduced by these static tokens. 
This  division of labor ensures a sophisticated functional specialization: \emph{early layers are "liberated" to comprehensively integrate cross-domain features}, while \emph{deep layers are entirely dedicated to localized temporal refinement}.

\begin{figure}[t]
\hspace{-2mm}
    \centering
    \includegraphics[width=\linewidth]{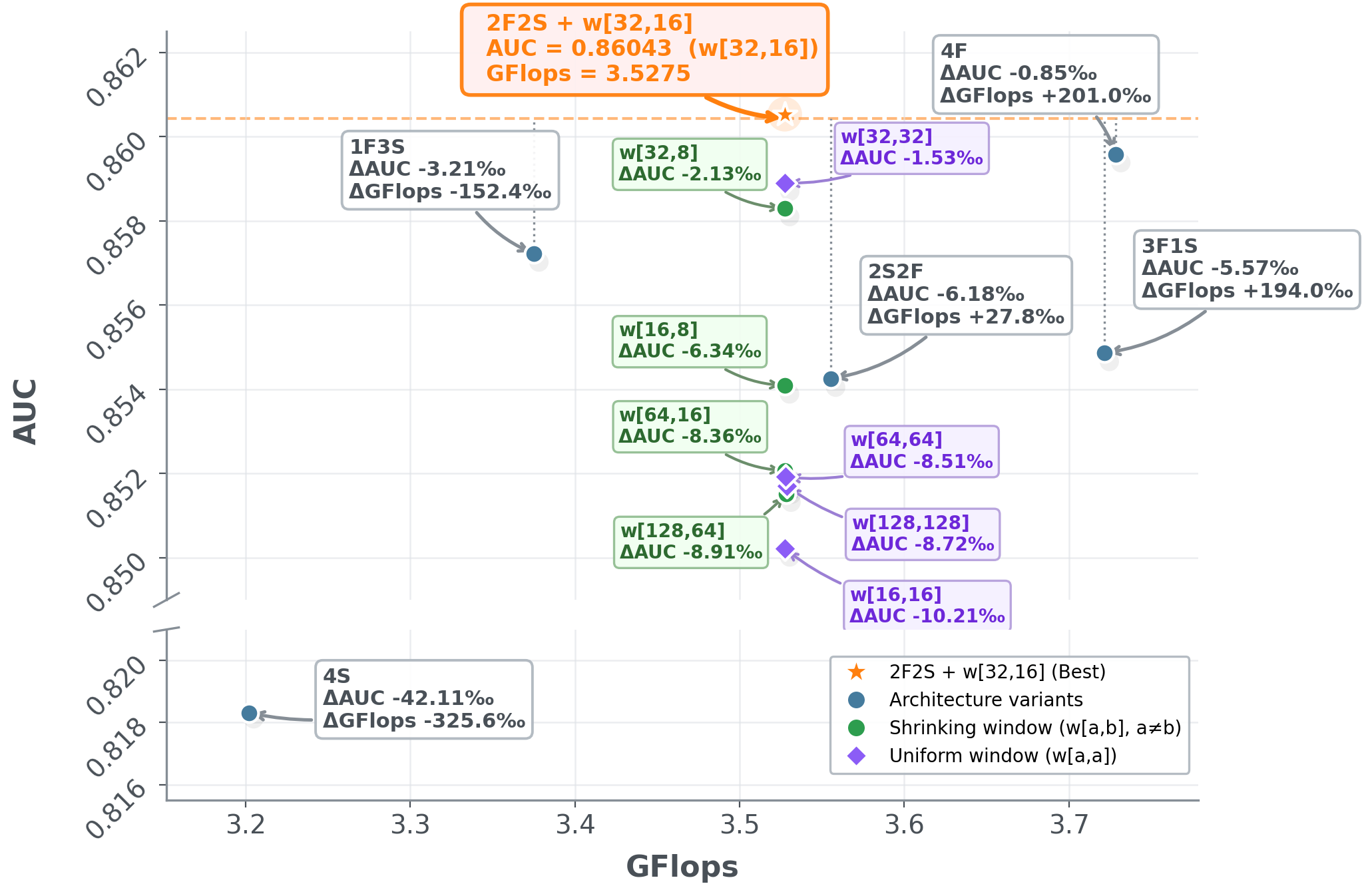}
    \caption{Efficiency and effectiveness trade-offs of various BFTS configurations. Explicit attention constraints of BFTS improve AUC, while linear-complexity SWA simultaneously reduces GFLOPs}
    \vspace{-2mm}
    \label{fig:gflops_auc}
\end{figure}

\subsection{Efficiency and Effectiveness of BFTS}
\label{sec:exp_bfts_dual_benefits}
In this section, we investigate whether the BFTS architecture can transcend the traditional trade-off between model performance and computational cost. Our core hypothesis is that while shallow layers require global visibility to fuse heterogeneous features, deeper layers benefit from localized attention, which filters out long-range noise and reduces complexity to near-linear $O(L)$.

The empirical results, as illustrated in Fig.~\ref{fig:gflops_auc}, reveal a significant performance gap tied to the hierarchical placement of attention. 
Our proposed {2F2S} architecture achieves the best AUC among all configurations, outperforming the {4F} baseline by 0.85‰ while reducing GFLOPs by 201.0‰. Reversing the order to {2S2F}, which prematurely constrains 
interactions in shallow layers, drops AUC by 6.18‰. 
These observations confirm that the architectural integration of BFTS delivers a dual benefit: it acts as a \emph{structural regularizer} that enhances representational purity while simultaneously \emph{lowering inference costs}.

Further sensitivity analysis indicates that the effectiveness of BFTS is closely tied to the sliding window size $w$. 
While overly expansive windows introduce distant noise and narrow windows truncate critical local dependencies, an optimal balance ($w=[32,16]$) yields the peak performance. 
This suggests that a properly calibrated window size serves as a vital inductive prior, ensuring the model focuses exclusively on the most pertinent behavioral dynamics in the final decision stages. Consequently, \textsc{TokenFormer} proves to be both a high-precision and high-efficiency backbone for large-scale recommendation.

Beyond the macro-architecture, we further hypothesize that the internal configuration of window sizes ($w$) should follow a progressively shrinking pattern across layers to facilitate hierarchical information distillation.
To verify this, we compare \textcolor{SYM}{uniform} and \textcolor{ASY}{shrinking window} configurations within the {2F2S} framework.
As summarized in Fig.~\ref{fig:gflops_auc}, a key observation emerges: \textcolor{ASY}{shrinking windows} consistently outperform their \textcolor{SYM}{uniform} counterparts. 
For instance, the configuration $w[32, 16]$ achieves the best AUC, surpassing the uniform $w[32, 32]$ by 1.53‰ in AUC. 
This progressively shrinking receptive-field bias effectively concentrates the model’s focus on the most immediate and pertinent user local temporal representation.

\begin{table}[t]
\centering
\caption{Architectural configurations of \textit{TokenFormer} across different model scales.}
\label{tab:model_scaling}
\begin{tabular}{lccc}
\toprule
\textbf{Model} & \textbf{Depth} & \textbf{Dim} & \textbf{Head Num} \\ 
\midrule
TokenFormer-T   & 4 & 64 & 4 \\
TokenFormer-S   & 4 & 256 & 4 \\
TokenFormer-M   & 6 & 256 & 4 \\
TokenFormer-L   & 8 & 256  & 8\\
\bottomrule
\end{tabular}
\end{table}

\subsection{Ablation Study}
\label{sec:exp_ablation}
To dissect the individual contributions of our proposed components and understand their synergistic effects, we conduct an ablation study on the KuaiRand-27k dataset. 
The results, summarized in Fig.\ref{fig:receptive_field_histograms_and_abl}, provide key insights into the efficacy of the Non-Linear Interaction Representation (NLIR) and the Bottom-Full-Top-Sliding (BFTS) strategy.

\textbf{Effectiveness of NLIR.} 
As reported in Fig.\ref{fig:receptive_field_histograms_and_abl}, incorporating \textcolor{GMcolor}{NLIR} alone yields a significant performance gain of \textbf{+4.87\textperthousand} AUC over the \textcolor{BSL}{Vanilla Transformer}. 
This improvement supports our hypothesis that standard linear attention often lacks the requisite complexity to capture higher-order feature correlations. 
By introducing explicit non-linear multiplicative interactions, NLIR not only enhances the \textit{expressiveness} of the learned embeddings but also contributes to \textit{dimensional robustness}. 
It mitigates the potential representation collapse when aligning sparse, low-dimensional features with dense, high-dimensional sequential tokens

\textbf{Impact of the BFTS Strategy.} 
A critical observation arises when applying Sliding Window Attention (SWA) across all layers (denoted as \textbf{4S}). 
This configuration leads to a catastrophic performance degradation, with the AUC dropping (\textbf{-36.35\textperthousand}). 
We attribute this sharp decline to the restricted receptive field inherent in a fully-windowed architecture, which inhibits the model's ability to capture long-range temporal dependencies and global sequential patterns, resulting in information fragmentation. 
In contrast, our \textcolor{BFTS}{BFTS strategy} mitigates this limitation by employing full attention in the early layers to establish comprehensive cross-token interactions, while reserving SWA for deeper layers to perform fine-grained, localized temporal refinement. 
As evidenced in Fig.\ref{fig:receptive_field_histograms_and_abl}, the BFTS strategy achieves a robust improvement of \textbf{+4.91\textperthousand} over the \textcolor{BSL}{Vanilla Transformer}, validating the necessity of hierarchical receptive field for modeling complex sequential patterns.

\textbf{Online Ablation Study.}
We further conduct an online ablation study in the Tencent Ads production environment using AUC as the evaluation metric. 
The production baseline is the conventional DLRM architecture, which has been incrementally trained on large-scale business data. 
Starting from a \textsc{TokenFormer} variant with \emph{full attention}, we observe a relative AUC change of \textbf{-0.16\%} compared with the production baseline. 
Replacing the full-attention backbone with the proposed \emph{BFTS mechanism} turns the result into a positive gain of \textbf{+0.14\%}. 
Finally, further equipping the BFTS backbone with the proposed \emph{NLIR} achieves the best performance, yielding a relative AUC improvement of \textbf{+0.22\%} over the production baseline.

\subsection{Model Scaling and Empirical Observations}
\label{sec:scaling_law}

To investigate the capacity potential of the \textit{TokenFormer} architecture, we conduct a scaling analysis by progressively increasing the model depth and hidden dimensions, as detailed in Table~\ref{tab:model_scaling}. 
Our empirical results on the public KuaiRand-27K dataset initially reveal a clear \textit{scaling law} phenomenon: as the model complexity expands from \textit{Tiny} to \textit{Large}, we observe consistent performance gains in ranking accuracy (see Table~\ref{tab:main_results}). 

However, this scaling trajectory encounters a plateau beyond the \textsc{TokenFormer-L} configuration. 
We attribute this saturation to the inherent \textit{data cardinality bottleneck} of the KuaiRand-27K dataset, where the limited sample volume proves insufficient to regularize a model of such high capacity, eventually leading to \emph{marginal overfitting}. 
In stark contrast, when deployed in our proprietary industrial production environment (\textit{Tencent Ads}), which operates at a significantly larger data scale, \textit{TokenFormer} continues to yield sustained performance improvements without exhibiting signs of premature saturation. 
This divergence suggests that while the architecture possesses a high theoretical upper bound for representation learning, its full potential is best unlocked in data-abundant regimes typical of large-scale industrial scenarios.

\subsection{Online A/B Tests}

To evaluate whether the offline gains of \textsc{TokenFormer} transfer to production, we deploy it in the WeChat Channels advertising system and conduct online A/B tests in the feed recommendation scenario. The experiment is carried out on real production traffic from January 2026 to February 2026, with the treatment model exposed to 5\% of the online traffic.

The online baseline follows a decoupled design for sequential and non-sequential features and is trained incrementally on continuously accumulated business data, making it a strong and highly optimized production system.
In contrast, \textsc{TokenFormer} replaces this separated modeling pipeline with a unified token-based architecture, where heterogeneous static fields and sequential behaviors are jointly modeled through the proposed BFTS and NLIR. \textsc{TokenFormer} is trained from scratch on two years of historical data before being deployed to the online A/B test.

We report \textbf{GMV} as the primary online business metric. Compared with the production baseline, \textsc{TokenFormer} achieves a \textbf{4.03\%} uplift in \textbf{GMV} during the A/B test. These results show that the gains observed offline can transfer to real serving conditions, and further confirm that the proposed unified architecture is practical and effective for large-scale industrial deployment.
\section{Conclusion}

In this paper, we identify Sequential Collapse Propagation (SCP) as a fundamental challenge in unifying multi-field and sequential recommendation. 
We empirically show that while non-sequential features provide informative priors, their low-rank nature can collapse behavioral representations within shared backbones. 
To mitigate this, we propose \textsc{TokenFormer}, integrating a bottom-full-top-sliding (BFTS) attention hierarchy and a non-linear interaction representation (NLIR) mechanism. 
Extensive experiments and online deployment in Tencent Ads demonstrate that \textsc{TokenFormer} not only achieves state-of-the-art accuracy but also recovers the intrinsic dimensionality of the representation manifold. 
Our work provides a robust blueprint for transitioning from heterogeneous expert ensembles toward unified, consistent recommendation backbones.

\bibliographystyle{ACM-Reference-Format}
\bibliography{references}

\appendix
\section{Complexity Analysis and Serving Optimization}
\label{app:complexity_analysis}

This appendix analyzes the computational complexity of the proposed \emph{Bottom-Full-Top-Sliding} (BFTS) design and the serving-side optimization used in online deployment. We focus on the dominant attention cost and omit lower-order terms such as normalization, bias addition, and point-wise nonlinearities.

\paragraph{Full attention vs. sliding-window attention.}
Consider an attention layer with sequence length $L$ and hidden dimension $d$. For full attention, both score computation and value aggregation are dominated by dense pairwise interactions, yielding
\begin{equation}
\mathcal{C}_{\mathrm{full}} = \mathcal{O}(L^2 d).
\label{eq:full_complexity}
\end{equation}
When sliding-window attention is used with window size $w$, each token attends to at most $w$ preceding tokens, reducing the complexity to
\begin{equation}
\mathcal{C}_{\mathrm{window}} = \mathcal{O}(L w d), \qquad w \ll L.
\label{eq:window_complexity}
\end{equation}
Hence,
\begin{equation}
\frac{\mathcal{C}_{\mathrm{window}}}{\mathcal{C}_{\mathrm{full}}}=\frac{w}{L},
\label{eq:complexity_ratio}
\end{equation}
showing that local attention becomes substantially cheaper when $w \ll L$.

\paragraph{Backbone complexity under BFTS.}
Suppose the \textsc{TokenFormer} backbone contains $L_f$ full-attention layers and $L_w$ sliding-window-attention layers. Its total complexity is
\begin{equation}
\mathcal{C}_{\mathrm{hybrid}}
=
\mathcal{O}\!\left(L_f L^2 d + L_w L w d\right),
\label{eq:hybrid_complexity}
\end{equation}
whereas an all-full-attention backbone of the same depth would require
\begin{equation}
\mathcal{C}_{\mathrm{all\text{-}full}}
=
\mathcal{O}\!\left((L_f+L_w)L^2 d\right).
\label{eq:allfull_complexity}
\end{equation}
Thus, BFTS preserves global modeling in lower layers while reducing the cost of upper layers through local attention. The same distinction also appears in memory usage:
\begin{equation}
\mathcal{M}_{\mathrm{full}} = \mathcal{O}(L^2), \qquad
\mathcal{M}_{\mathrm{window}} = \mathcal{O}(Lw).
\end{equation}

\paragraph{Serving complexity under joint encoding.}
Let $L_u$ and $L_a$ denote the numbers of user-side and ad-side tokens, respectively, and let $B$ be the number of candidate ads scored for one request. Under a straightforward joint-encoding strategy, each candidate is scored with a concatenated sequence of length $L_u+L_a$, giving
\begin{equation}
\mathcal{C}_{\mathrm{serve}}^{\mathrm{joint}}
=
\mathcal{O}\!\left(B(L_u+L_a)^2 d\right).
\label{eq:joint_serving_complexity}
\end{equation}
A key inefficiency here is that the user-side self-interaction is recomputed for every candidate, even though the user context is shared.

\paragraph{Serving complexity under decoupled encoding.}
To remove this redundancy, we adopt a decoupled serving strategy. The user-side tokens are encoded once and compressed into $N$ summary tokens, which are then combined with each candidate ad for scoring. The resulting complexity is
\begin{equation}
\mathcal{C}_{\mathrm{serve}}^{\mathrm{decouple}}
=
\mathcal{O}\!\left(L_u^2 d + B(N+L_a)^2 d\right).
\label{eq:decoupled_serving_complexity}
\end{equation}
Compared with Eq.~\ref{eq:joint_serving_complexity}, the quadratic user-side term is no longer multiplied by $B$, and the candidate-specific cost depends on the compressed representation length $N$ rather than the original user sequence length $L_u$. The complexity gap between the two serving strategies can be written as
\begin{equation}
\Delta \mathcal{C}
=
\mathcal{C}_{\mathrm{serve}}^{\mathrm{joint}}
-
\mathcal{C}_{\mathrm{serve}}^{\mathrm{decouple}}
\approx
\mathcal{O}\!\left((B-1)L_u^2 d + B\big[(L_u+L_a)^2-(N+L_a)^2\big]d\right).
\label{eq:serving_gap}
\end{equation}
As long as $B>1$ and $N<L_u$, the decoupled design is asymptotically more efficient.

\paragraph{Practical serving improvement.}
In our online deployment, the decoupled inference strategy improves serving throughput from 126 QPS to 695 QPS, corresponding to a 5.5$\times$ speedup. This gain is consistent with the above analysis: it comes from removing repeated user-side computation and combining it with the architectural sparsification induced by BFTS.

\paragraph{Summary.}
Overall, \textsc{TokenFormer} improves efficiency from two complementary perspectives: BFTS reduces the cost of upper layers through local attention, while decoupled serving amortizes user-side computation across candidate ads. Together, these two mechanisms make the model substantially more practical for large-scale industrial deployment.

\section{Analysis Tools}
\subsection{Mutual Information Diagnostics of Hidden Representations}
\label{app:mi_diagnostics}

This appendix details the computation of the mutual information (MI) diagnostics used in Section~\ref{sec:exp_gated_dimension}. These diagnostics quantify how much task-relevant discriminative information is retained in the model's output representations with respect to each user action.

\paragraph{Representation extraction.}
For each model variant, we perform a forward pass over the test set and extract three types of representations: (i)~the raw logit output $\mathbf{X}_{\mathrm{raw}} \in \mathbb{R}^{N \times d}$, (ii)~the sigmoid-activated output $\mathbf{X}_{\mathrm{sig}} \in \mathbb{R}^{N \times d}$, and (iii)~the penultimate-layer hidden states $\mathbf{X}_{\mathrm{pen}} \in \mathbb{R}^{N \times d_h}$, where $N$ is the number of test samples and $d$, $d_h$ denote the output and hidden dimensions, respectively.

\paragraph{Multi-label construction.}
Our recommendation task involves multiple user actions. For each sample~$i$, we extract a binary label vector
\begin{equation}
    \mathbf{y}_i = \bigl(y_i^{(1)}, \dots, y_i^{(A)}\bigr) \in \{0,1\}^{A},
\end{equation}
where $A$ denotes the number of action types (\emph{e.g.}, click, like, follow, comment, forward) and $y_i^{(a)} = 1$ indicates that the user performed action~$a$ on the target item. Each action is treated as an independent binary classification task for MI estimation.

\paragraph{Continuous MI estimation.}
We estimate the mutual information between the continuous representations and each binary action label using the $k$-nearest-neighbor-based KSG estimator. For each action~$a$, the per-feature MI scores are aggregated to obtain a scalar:
\begin{equation}
    \mathrm{MI}_{\mathrm{KSG}}^{(a)}
    = \sum_{j=1}^{d} \widehat{I}\!\left(X_j;\, Y^{(a)}\right),
\end{equation}
where $\widehat{I}(\cdot;\cdot)$ denotes the KSG estimate for the $j$-th feature dimension and the $a$-th action label. The KSG estimator operates directly on continuous representations, thereby avoiding information loss from discretization.

\paragraph{Discretized MI via KMeans clustering.}
As a complementary approach, we discretize the continuous representations using KMeans clustering. For a given number of clusters~$K$, we fit KMeans on $\mathbf{X}_{\mathrm{raw}}$ and assign each sample to its nearest cluster center:
\begin{equation}
    c_i^{(K)} = \mathrm{KMeans}\!\left(\mathbf{x}_i;\, K\right), \qquad i = 1, \dots, N.
\end{equation}
We then compute the discrete mutual information between the cluster variable~$C$ and each action label~$Y^{(a)}$:
\begin{equation}
    \mathrm{MI}^{(a)}(K)
    =
    \sum_{c \in \mathcal{C}} \sum_{y \in \{0,1\}}
    p(c,y)\log\frac{p(c,y)}{p(c)\,p(y)},
\end{equation}
where $\mathcal{C}=\{1,\dots,K\}$. We evaluate the following cluster numbers:
\begin{equation}
    K \in \{4, 8, 16, 32, 48, 64, 96\},
\end{equation}
to assess the robustness of the MI estimates with respect to discretization granularity.

\paragraph{Weighted aggregation across actions.}
Since different actions exhibit vastly different positive rates, we aggregate per-action MI scores into a single summary statistic via a weighted mean:
\begin{equation}
    \overline{\mathrm{MI}} = \frac{\sum_{a=1}^{A} w_a \cdot \mathrm{MI}^{(a)}}{\sum_{a=1}^{A} w_a},
\end{equation}
where the weights $w_a$ reflect the relative importance of each action. When not explicitly specified, the weights default to the positive sample counts $w_a = n_a^{+} = \sum_i y_i^{(a)}$, ensuring that rare but important actions are not dominated by high-frequency ones.

\paragraph{Multi-seed aggregation.}
To reduce variance from random initialization, we repeat all experiments across multiple random seeds. For each model variant, we compute MI diagnostics independently per seed and report the mean across seeds.

\paragraph{Interpretation.}
The MI diagnostics provide a functional assessment of representation quality by directly measuring how much discriminative information the learned representations preserve for each downstream action. The continuous KSG estimator avoids discretization artifacts, while the KMeans-based MI offers interpretability through explicit clustering structure. Together, they provide complementary views of the information content from both nonparametric and partition-based perspectives.

\subsection{Spectral Diagnostics of Hidden Representations}
\label{app:spectral_diagnostics}

\begin{figure}[t]
\vspace{-2mm}
    \centering
    \includegraphics[width=\linewidth]{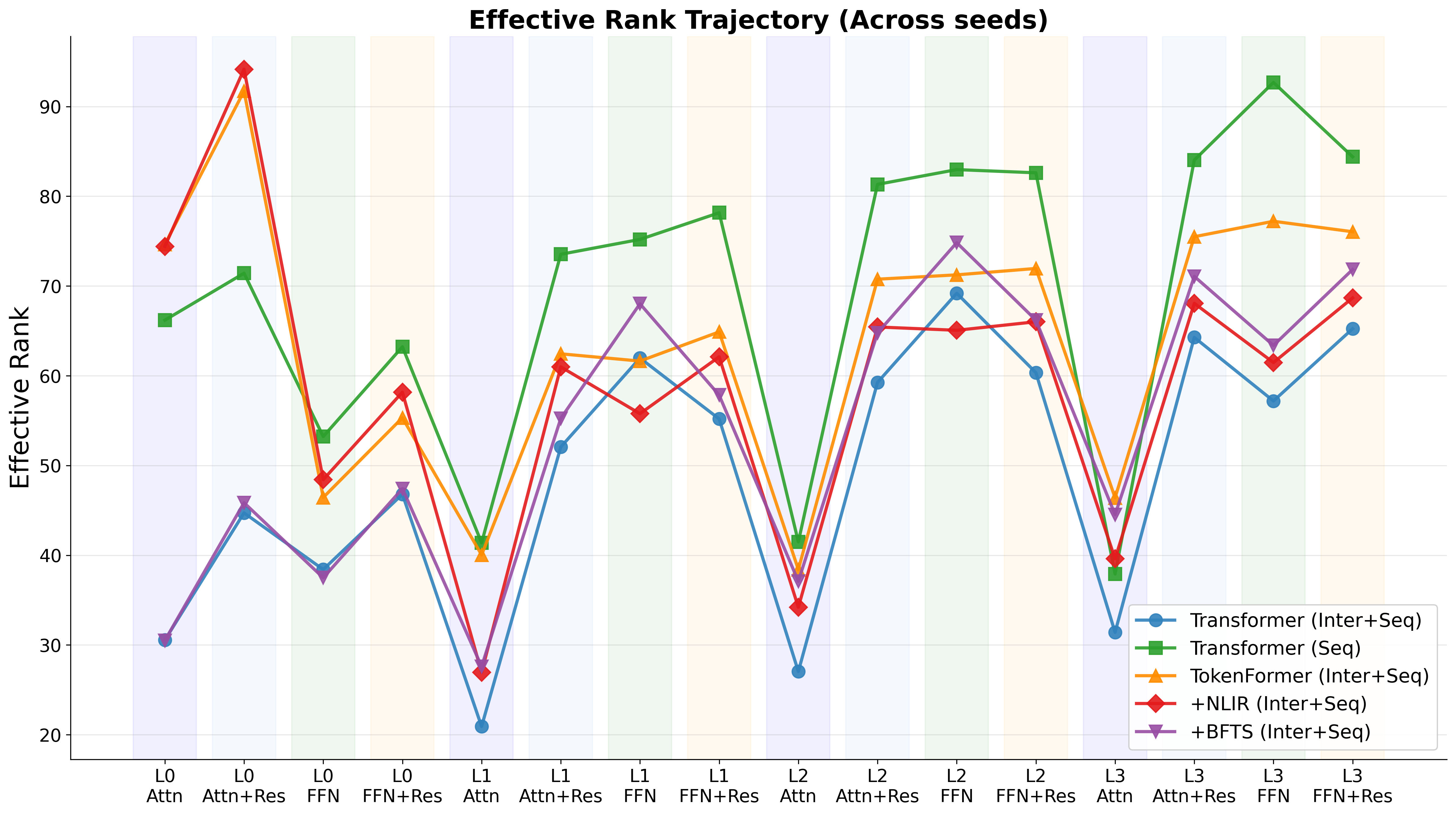}
    \vspace{-2mm}
    \caption{Block-wise effective-rank trajectory of sequential behavioral tokens ($\Tokens$) for the \textcolor{BSL}{Vanilla Transformer}, \textcolor{GMcolor}{Transformer} with only sequence tokens, \textcolor{TKF}{\textsc{TokenFormer}}, \textcolor{TEASER_BFTS}{$+$BFTS} and \textcolor{TEASER_GM}{$+$NLIR}. 
    We track the output across each block include: Attention output (Eq.\ref{eq:ugi-attn}), the successive residual additions (Eq.\ref{eq:attn-residual}), and the FFN output (Eq.\ref{eq:swiglu-rmsnorm}).    }
    \vspace{-2mm}
    \label{fig:norm_sv}
\end{figure}

This appendix details the computation of the spectral diagnostics used in Section~\ref{sec:exp_gated_dimension}, including the effective rank and the normalized singular value spectrum.

\paragraph{Hidden representation extraction.}
For each layer $l$, we collect hidden representations from a forward pass over the test set. We retain only the item and candidate-item tokens, identified by their token types, and stack their hidden states into a representation matrix
\begin{equation}
    \mathbf{X}^{(l)} \in \mathbb{R}^{S_l \times d},
\end{equation}
where $S_l$ denotes the number of collected tokens at layer $l$, and $d$ is the hidden dimension. In our experiments, $d=256$. To keep the analysis computationally tractable, we cap the number of sampled tokens at $S_{\max}=10{,}000$.

\paragraph{Singular value decomposition.}
Given the representation matrix $\mathbf{X}^{(l)}$, we first center it by subtracting the column-wise mean:
\begin{equation}
    \bar{\mathbf{X}}^{(l)} = \mathbf{X}^{(l)} - \mathbf{1}_{S_l} \boldsymbol{\mu}^{(l)\top}, 
    \qquad 
    \boldsymbol{\mu}^{(l)} = \frac{1}{S_l}\sum_{i=1}^{S_l} \mathbf{x}_i^{(l)},
\end{equation}
where $\mathbf{x}_i^{(l)}$ denotes the $i$-th row of $\mathbf{X}^{(l)}$. We then perform singular value decomposition (SVD):
\begin{equation}
    \bar{\mathbf{X}}^{(l)} = \mathbf{U}^{(l)} \mathbf{\Sigma}^{(l)} {\mathbf{V}^{(l)}}^\top,
\end{equation}
with
\begin{equation}
    \mathbf{\Sigma}^{(l)} = \mathrm{diag}\!\left(s_1^{(l)}, s_2^{(l)}, \dots, s_d^{(l)}\right),
\end{equation}
where $\{s_k^{(l)}\}_{k=1}^{d}$ are the singular values sorted in descending order.

\paragraph{Effective rank.}
To quantify subspace utilization, we compute the entropy-based effective rank from the singular value distribution. Specifically, we first normalize the singular values as
\begin{equation}
    p_k^{(l)} = \frac{s_k^{(l)}}{\sum_{j=1}^{d} s_j^{(l)}}, 
    \qquad 
    k=1,\dots,d.
\end{equation}
The effective rank of layer $l$ is then defined as
\begin{equation}
    r_{\mathrm{eff}}^{(l)} 
    = \exp\!\left(
    -\sum_{k=1}^{d} p_k^{(l)} \log p_k^{(l)}
    \right).
    \label{eq:appendix_effective_rank}
\end{equation}
A larger effective rank indicates that variance is distributed across a broader set of latent dimensions, whereas a smaller value implies stronger concentration in a low-dimensional subspace.

\paragraph{Normalized singular value spectrum.}
To characterize how variance is distributed across latent dimensions, we further compute the normalized singular value spectrum by dividing each singular value by the largest one:
\begin{equation}
    \tilde{s}_k^{(l)} = \frac{s_k^{(l)}}{s_1^{(l)}}, 
    \qquad 
    k=1,\dots,d.
    \label{eq:appendix_norm_sv}
\end{equation}
The resulting spectrum $\{\tilde{s}_k^{(l)}\}_{k=1}^{d}$ satisfies $\tilde{s}_1^{(l)}=1$ and is monotonically non-increasing. A slower decay indicates that variance is spread more evenly across dimensions, while a steeper decay suggests that the representation is dominated by a small number of leading singular directions.

\paragraph{Layer-wise and stage-wise analysis.}
For the layer-wise analysis, we compute the above diagnostics independently for each layer $l$. For the intra-block analysis in Section~\ref{sec:exp_gated_dimension}, the same procedure is applied to hidden representations collected at three stages within each Transformer block: (i) the attention output before the residual connection, (ii) the MLP/FFN output before the residual connection, and (iii) the final block output after residual addition.

\paragraph{Interpretation.}
The spectral diagnostics capture complementary aspects of representation geometry. Effective rank reflects the overall degree of subspace utilization, whereas the normalized singular value spectrum reveals how variance is distributed among singular directions. Together, they provide a compact characterization of dimensional robustness in the learned hidden representations.

\section{Normalized Singular Value Spectrum Analysis}
\label{app:norm_sv_analysis}

This appendix provides a complementary spectral analysis of hidden representations using normalized singular value spectra.

\begin{figure}[t]
\vspace{-2mm}
    \centering
    \includegraphics[width=\linewidth]{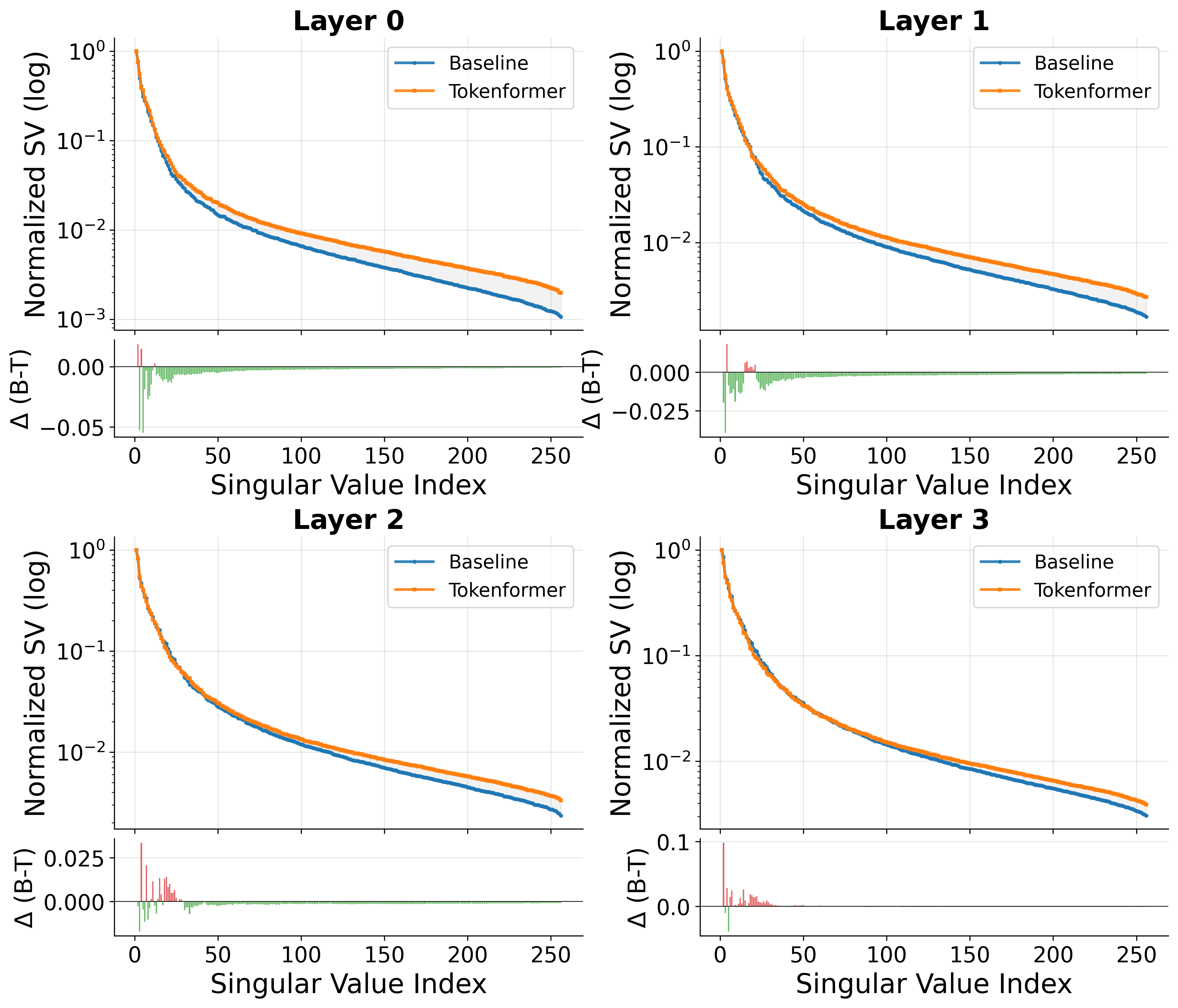}
    \vspace{-2mm}
    \caption{Per-layer normalized singular value spectra. Compared with the non-gated baseline, \textsc{TokenFormer} exhibits less top-heavy spectra, indicating a more balanced allocation of variance across latent dimensions.}
    \vspace{-2mm}
    \label{fig:norm_sv}
\end{figure}

To further examine how variance is distributed across latent dimensions, we compare the normalized singular value spectra of the gated and non-gated models. As shown in Figure~\ref{fig:norm_sv}, across all layers, the gated \textcolor{TKF}{\textsc{TokenFormer}} exhibits a less top-heavy spectrum and a heavier tail than the \textcolor{BSL}{non-gated variant}. In other words, variance is distributed more evenly across dimensions rather than being dominated by a few singular directions.

This result provides complementary evidence for the same conclusion as the effective-rank analysis. While the effective rank reflects the overall degree of subspace utilization, the spectral distribution reveals how variance is structurally allocated within that subspace. Taken together, both observations support the claim that explicit gating improves dimensional robustness by preserving richer and more balanced representations. This also provides a geometric explanation for the downstream ranking gains of \textsc{TokenFormer}: better-preserved representation diversity leads to stronger and more stable sequence modeling.
\section{Discussion}

\subsection{RoPE-based Positional Awareness}
\label{sec:hybrid-masking}

\noindent \textbf{From Additive to Multiplicative Dynamics.}
Contrary to conventional recommendation paradigms that often marginalize positional embeddings arguing that explicit timestamps in side-information provide sufficient \textit{implicit chronology}.
We argue that capturing fine-grained relative dependencies is pivotal for modeling user behavior evolution. 
Standard architectures typically employ additive positional encodings, where the attention score between a query $\mathbf{q}$ at position $m$ and a key $\mathbf{k}$ at position $n$ is:
\begin{equation}
    \text{Attn}(m, n) = (\mathbf{q} + \mathbf{p}_m)^\top (\mathbf{k} + \mathbf{p}_n) = \underbrace{\mathbf{q}^\top \mathbf{k}}_{\text{semantic}} + \underbrace{\mathbf{q}^\top \mathbf{p}_n + \mathbf{p}_m^\top \mathbf{k}}_{\text{cross-terms}} + \underbrace{\mathbf{p}_m^\top \mathbf{p}_n}_{\text{bias}},
\end{equation}
where $\mathbf{p}$ denotes absolute positional embeddings. 
We observe that such additive schemes inevitably introduce detrimental noise through the cross-terms. 
Because positional embeddings typically reside in a lower-intrinsic-dimensional subspace, forcing additive interactions between high-dimensional semantic states and low-rank positional vectors can trigger severe \textit{representation interference}, 
where the intrinsic geometry of the semantic manifold is compromised to satisfy rigid positional constraints.

To preserve representation integrity, \textsc{TokenFormer} adopts multiplicative \textbf{RoPE}. 
By rotating hidden states in the complex plane, the interaction is redefined as:
\begin{equation}
    \text{Attn}(m, n) = (\mathbf{R}_m \mathbf{q})^\top (\mathbf{R}_n \mathbf{k}) = \mathbf{q}^\top (\mathbf{R}_m^\top \mathbf{R}_n) \mathbf{k} = \mathbf{q}^\top \mathbf{R}_{n-m} \mathbf{k}.
\end{equation}
Crucially, unlike additive projections, $\mathbf{R}_{n-m}$ is an orthogonal matrix. This unitary transformation acts as a full-rank isometry, strictly preserving the $L_2$ norm and the intrinsic dimensionality of the semantic representations without squashing them into a low-rank subspace. 
This formulation elegantly encodes the relative distance $n-m$ directly into the phase of the hidden states. 
Therefore, \textsc{TokenFormer} completely circumvents the risk of rank collapse, maintaining the full expressive power of the semantic manifold while achieving superior length extrapolation.

\noindent \textbf{Type-Aware Feature Interactions.}
According to the properties of RoPE, the relative rotation matrix between any two static tokens is $\mathbf{R}_{0-0} = \mathbf{I}$. 
Consequently, the attention score between field tokens degenerates into a pure semantic dot-product: $\text{Attn}(0, 0) = \mathbf{q}^\top \mathbf{k}$.

By sharing position $p_0$, the Transformer backbone executes multi-order feature interactions analogous to a Factorization Machine (FM) without positional interference. 
Furthermore, when sequential tokens interact with static fields, the relative rotation $\mathbf{R}_{n-0}$ allows the model to dynamically modulate its reliance on static profiles based on the specific temporal stage $n$ of the user's journey, learning how historical context should be weighted against long-term user preferences.

\subsection{Gated Mechanism}
While gated mechanisms are broadly adopted~\cite{huang2019fibinet, PEPNet, AFN, qiu2025gatedattn}, we specifically theoretically frame their necessity in our unified architecture through two critical lenses:

\noindent \textbf{Representation Expressiveness.} Capturing sparse ID co-occur-rences inherently requires explicit multiplicative interactions~\cite{rendle2020neural, zhai2024hstu}, which vanilla attention lacks during value aggregation. 
The gating formulation (i.e., Eq.\ref{eq:ugi-gate}) injects this requisite nonlinearity, expanding the model's hypothesis space (detailed in Sec.~\ref{sec:exp_gated_expressiveness}).

\noindent \textbf{Dimensional Robustness.} According to Interaction Collapse Theory~\cite{guo2024embeddingcollapse}, unifying low-cardinality static features with dynamic sequences amplifies the risk of rank collapse. 
The nonlinear multiplicative operator theoretically preserves the dimensional rank of $\mathbf{X}_1$, effectively truncating the collapse propagation across domains~\cite{yin2025feature} (empirically validated in Sec.~\ref{sec:exp_gated_dimension}).

\end{document}